\documentclass[12pt]{article}
\pdfoutput=1
\usepackage{jheppub}
\usepackage{amsmath}
\usepackage{MnSymbol}
\usepackage{wrapfig}
\usepackage[rightcaption]{sidecap}
\usepackage{graphicx} 
\graphicspath{ {images/} }

\usepackage{tikz}
\tikzset{>=latex}
\tikzset{baseline=(A.base)}
\tikzset{every picture/.append style={scale=1.1}}
\tikzset{flavor/.style={draw}}
\tikzset{gauge/.style={draw,circle,inner sep=2pt}}

\def\be{\begin{equation}}
\def\ee{\end{equation}}
\def\ba{\begin{aligned}}
\def\ea{\end{aligned}}
\def\ben{\begin{eqnarray}\displaystyle}
\def\een{\end{eqnarray}}

\def\CW{{\cal W}}
\def\CF{{\cal F}}
\def\CL{{\cal L}}
\def\CT{{\cal T}}
\def\HS{{\cal HS}}
\def\CN{{\cal N}}
\def\IZ{{\mathbb Z}}
\def\M{{\mathfrak{M}}}

\def\At{{\tilde A}}
\def\Bt{{\tilde B}}
\def\Ct{{\tilde C}}

\def\CLt{{\tilde \CL}}
\def\Mt{{\tilde M}}
\def\at{{\tilde a}}
\def\bt{{\tilde b}}
\def\ct{{\tilde c}}
\def\pt{{\tilde p}}
\def\Pt{{\tilde P}}
\def\qt{{\tilde q}}
\def\Qt{{\tilde Q}}

\def\nn{\nonumber}

\preprint{
{\small{\textsf{}}}}

\title{3d $\CN=2$ mirror symmetry, pq-webs\\ and monopole superpotentials}

\author[1]{Sergio Benvenuti}
\author[2]{and Sara Pasquetti}

\affiliation[1]{International School of Advanced Studies (SISSA),\\
via Bonomea 265, 34136 Trieste, Italy and INFN, Sezione di Trieste.}
\affiliation[2]{Dipartimento di Fisica, Universit\`a di Milano-Bicocca, I-20126 Milano, Italy}
\emailAdd{benve79@gmail.com, sara.pasquetti@gmail.com.}

\abstract{D3 branes stretching between webs of (p,q) 5branes provide an interesting class of $3d$ $\CN=2$ theories.
For generic pq-webs however the low energy field theory is not known. 
We use 3d mirror symmetry and Type IIB S-duality to construct Abelian gauge theories corresponding to  D3 branes ending on both sides of a pq-web made of many coincident $NS5$'s intersecting one $D5$. These theories contain chiral monopole operators in the superpotential and enjoy a non trivial pattern of global symmetry enhancements. In the special case of the pq-web with one $D5$ and one $NS5$, the 3d low energy SCFT admits three dual formulations. This triality  can be applied locally inside bigger quiver gauge theories. We prove our statements using partial mirror symmetry \`a la Kapustin-Strassler, showing the equality of the $S^3_b$ partition functions and studying the quantum chiral rings. 
}

\begin{document}

\maketitle

\section{Introduction and summary}

Two decades ago Intriligator and Seiberg \cite{Intriligator:1996ex} discovered three dimensional mirror symmetry: a quantum duality acting on theories with $8$ supercharges. One salient feature of this duality is the exchange of monopole operators with standard mesonic operators, or of Coulomb branches with Higgs branches. Soon after, 3d mirror symmetry was interpreted as Type IIB S-duality acting on $1/4$ BPS brane setups composed of $D3$, $D5$ and $NS5$ branes \cite{Hanany:1996ie}. These Hanany-Witten setups can be easily generalized to field theories living in a different number of dimensions.

Extending the analysis to models with only $4$ supercharges is clearly desirable, but making progress proved quite difficult. $3d$ $\CN=2$ mirror symmetry is understood in the Abelian case  \cite{Aharony:1997bx, deBoer:1997kr}. Two important results using brane setups  are the brane interpretation of Seiberg dualities and the construction of chiral field theories \cite{Aharony:1997ju, Brunner:1998jr} (with $SU(N_f) \times SU(N_f)$ chiral global symmetry) from $Dp$ branes ending on $D(p+2)$-$NS5$ branes intersections. In the $3d$ case we have $D3$ branes ending on pq-webs \cite{Aharony:1997ju,Aharony:1997bh},  $1/4$ BPS brane setups involving $D5'$ and $NS5$ branes that give rise to five dimensional SCFT's with minimal supersymmetry.

Our aim is to tackle some problems that to the best of our knowledge are still open. We focus on the set-up consisting of a single $D3$ brane ending on both sides of a pq-web composed of $1$ $NS5$ brane and $K$ $D5'$ branes, the $(1_{NS5} ,  K_{D5'})$ web, and on the S-dual  $(K_{NS5} ,  1_{D5'})$ web. We also terminate the $D3$ branes on two spectator $NS5'/D5'$ branes.
The low energy theory corresponding to the $(1_{NS5} ,  K_{D5'})$ set-up is known, it is the $U(1) \times U(1)$ quiver with $K$ flavors for each node \cite{Aharony:1997ju, Brunner:1998jr} depicted in fig. \ref{tetrak0}, with the so-called {\it flavor doubling}: the $D5'$ branes are broken into two pieces and each half provides a $SU(K)$ mesonic symmetry. The two sets of flavors {\it talk} to each other through a cubic superpotential that preserves a $SU(K) \times SU(K)$ chiral global symmetry. 

The low description of the S-dual set-up was not known and we derive it by using partial Abelian mirror symmetry \`a la Kapustin-Strassler \cite{Kapustin:1999ha}, it is the quiver theory composed of two $U(1)^{K-1}$ tails sketched in  fig. \ref{tetrak0}. Each tails supports an $U(1)_{top}^{K-1}$ topological symmetry and a $U(1)_{axial}^{K-1}$ axial symmetry which  enhances to $SU(K)_{top} \times SU(K)_{axial}$. The two tails {\it talk} to each other through a superpotential which is a sum of monopole operators involving gauge groups on both tails. This superpotential has the non-trivial effect 
of breaking the total enhanced global symmetry to the diagonal $SU(K)_{top} \times SU(K)_{axial}$. We then see that the mirror theory shows  {\it gauge doubling}: the $NS5$ branes are broken into two pieces and each half provides a $SU(K)$ topological/axial symmetry.
 
Our main tools to complement our analysis via partial mirror symmetry are the study of three-sphere partition functions
and the analysis of  the chiral rings.

 The method of localisation applied to SUSY theories defined on compact spaces,  pioneered by Pestun in the case of 4d $\mathcal{N}=2$ theories on $S^4$ \cite{Pestun:2007rz} and then widely generalized, allows us to obtain exact results for partition functions and other observables. Path integrals reduce to ordinary matrix integrals which are explicit  functions of global symmetry fugacities. Testing dualities amounts to show the equivalence of the matrix integrals under an appropriate map of the fugacities.

To study the chiral rings of mirror theories we rely on the recent progress in  understanding of monopole operators in $3d$ $\CN=4$ and $\CN=2$ gauge theories \cite{Borokhov:2002ib,Borokhov:2002cg,Gaiotto:2008ak,Benna:2009xd, Bashkirov:2010kz, Imamura:2011su, Cremonesi:2013lqa, Cremonesi:2014kwa, Cremonesi:2014vla}.
Having a good control on the monopole operators is crucial in our case since these operators appear  in the superpotentials of  our theories.

 Polyakov showed long time ago that monopole operators enter the effective potential in field theories arising at low energies from spontaneous symmetry breaking \cite{Polyakov:1976fu, Affleck:1982as}. It is also known that compactifying $4d$ supersymmetric field theories on a circle introduces monopole operators in the low energy $3d$ superpotential \cite{Seiberg:1996nz, Aharony:2013dha}.

Actually  since the general paradigm is that  mirror symmetry acts on the chiral rings
mapping mesonic operators into monopole operators, whenever  we have generic  mesonic operators in the superpotential we  should expect  monopole operators in the superpotential of the dual theory. We will see many examples of this.
\footnote{See also \cite{Collinucci:2016hpz} for an example of monopole operators in the superpotential.}

 In section \ref{prel} we review the Abelian mirror between $U(1)$ with $N_f$ flavors and the linear quiver $U(1)^{N_f-1}$. We detail a proof of the duality, using partial mirror symmetry, and present a complete map between the two quantum chiral rings.
  
 In section \ref{triality} we discuss in detail the pq-web $(1_{NS5} ,  1_{D5'})$, which under S-duality is mapped to itself. We find a proof that the gauge theory associated to the brane setup $1_{NS5'} - 1_{D3} - (1_{NS5} ,  1_{D5'}) - 1_{D3} - 1_{NS5'}$ is dual of the Wess-Zumino $XYZ+X'Y'Z'$ model associated to the S-dual setup $1_{D5'} - 1_{D3} - (1_{NS5} ,  1_{D5'}) - 1_{D3} - 1_{D5'}$.

 \begin{wrapfigure}{r}{0.43\textwidth}
\centering
\includegraphics[
width=0.47\textwidth]{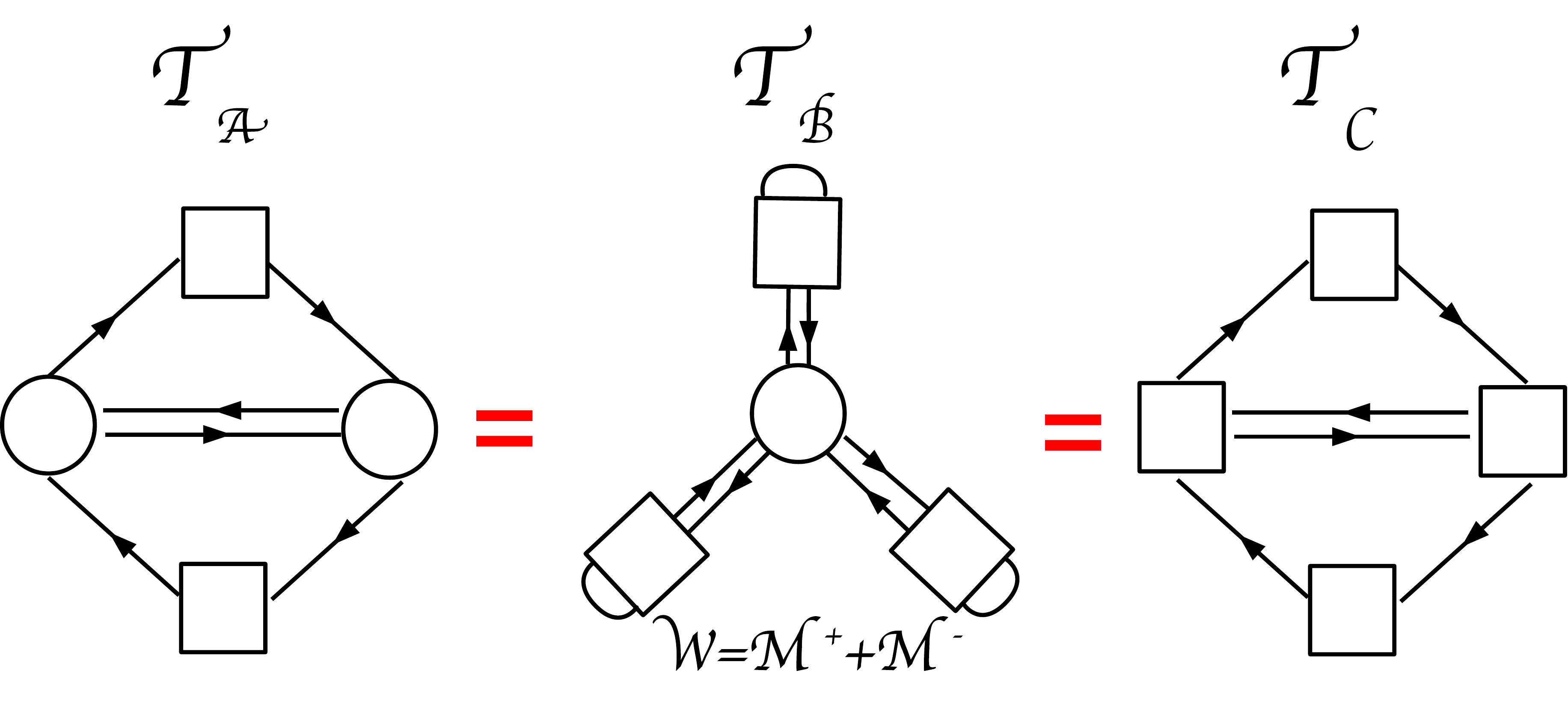}
\caption{The gauge theory triality. Circles stand for $U(1)$ gauge groups, squares stand for global $U(1)$ symmetries. In the central quiver part of the superpotential is the sum of the two basic monopole operators for the $U(1)$ node.}
\label{triweb0}
\end{wrapfigure}
 Along the way we discover a triality with a third theory: a $U(1)$ gauge theory with $3$ flavors and a superpotential containing the two basic monopole operators $\CW= \M^+ + \M^-$. The duality, when projected down to the equality of the corresponding $S^3_b$ partition functions, becomes an integral identity which appeared in the mathematical literature 
 as the {\it new pentagon} or {\it ultimate integral identity}. Here we provide a brane realization and a field theory interpretation of that identity and   prove the duality at the level of the full gauge theory. A similar relation for $U(1)$ with $4$ flavors and monopole operators in the superpotential has been discussed in \cite{Dimofte:2012pd} and plays an important role in the 3d-3d correspondence.

We then generalize to our main result of $D3$ branes ending on the $(K_{D5} ,  1_{NS5'}) \leftrightarrow (K_{NS5} ,  1_{D5'})$ pq-webs in section \ref{k1web}. We also show that the duality discussed above between $U(1)$ with $3$ flavors and the $XYZ+X'Y'Z'$ model can be used to provide two different presentation of the theories (two on each side of the duality).

\begin{SCfigure}[1.9][h]
\centering
\includegraphics[height=9cm,width=0.7\textwidth]{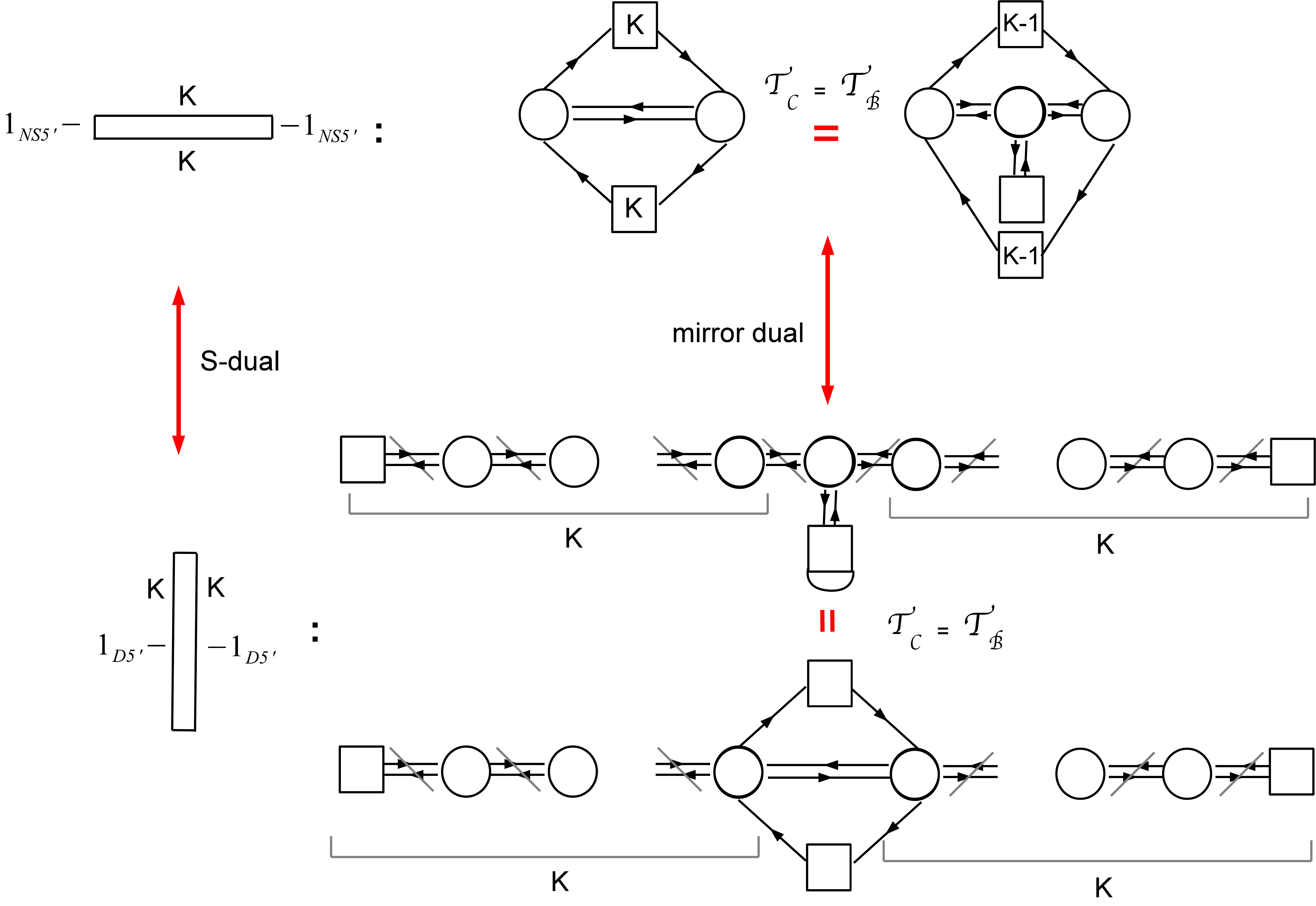}
\caption{Duality patterns for  gauge theories associated to S-dual brane setups. The quivers in the first line are related by the application
of the duality  $\CT_B \leftrightarrow \CT_C$ of figure \ref{triweb0} and describe the $(1_{NS5}, K_{D5'})$-web.  
The quivers in the third and forth lines contain many monopole terms in the superpotential, are related by the application
of the duality  $\CT_B \leftrightarrow \CT_C$ and describe $D3$ branes ending on the S-dual $(K_{NS5}, 1_{D5'})$-web.}
\label{tetrak0}
\end{SCfigure}

Notice that the symmetry enhancement pattern $U(1)^{k-1} \times U(1)^{k-1} \rightarrow SU(k) \times SU(k)$  of the theory in the last line of fig. \ref{tetrak0} is quite peculiar: on each side one node has $3$ flavors, while the other $k-2$ nodes have $2$ flavors. Nevertheless the monopoles involving the $3$-flavors-node combine with the other monopoles to form a bifundamental representation of $SU(k) \times SU(k)$. This shows that the topological symmetry enhancement in this $\CN=2$ case follows rules that are different from the  $\CN=4$ case, where only sub quivers with balanced $U(N)$ nodes (number of flavors equal to  twice the number of colors)  are expected to have an enhanced topological symmetry. It would be interesting to investigate these issues further.

In section \ref{deform} we study relevant deformations of the dualities. Real mass deformations lead, on the mirror side, to chiral theories with Chern-Simons interactions, both for the same gauge groups (CS couplings) and for different gauge groups (BF couplings).

In section \ref{necklace} we show how similar techniques can be applied to longer quivers. We consider a circular quiver with flavors at each node that is self-mirror.


\section{Preliminaries: Abelian mirror symmetry, $S^3$ partition functions and quantum chiral rings}\label{prel}
In this section we review some basic examples of  Abelian mirror symmetry for 3d $\mathcal{N}=2$ theories \cite{Aharony:1997bx}\cite{Intriligator:1996ex}\cite{deBoer:1997kr} focusing on the  partial mirror symmetry approach  \`a  la Kapustin-Strassler \cite{Kapustin:1999ha}. We complement this derivation with the  study of  the mapping of the quantum chiral rings of mirror theories.

We begin with the $\mathcal{N}=4$ case.
The most basic example is the duality between the SQED  with one hyper and the theory of a free (twisted) hyper.
In \cite{Borokhov:2002cg} an exact proof of this duality was given, showing that the monopole operators $\M^\pm$ of the SQED are free fields with scaling dimension $\Delta=1/2$, hence equivalent to a free hyper.

In \cite{Kapustin:1999ha} it was shown that all the  Abelian mirror duals can be derived from this basic Abelian duality.
The key idea is to regard partition functions as functionals of  background vector multiplets $\hat V$ associated to the global symmetries of the theory and to prove that the basic Abelian mirror duality is in fact an invariance under Functional Fourier Transform.

The  partition function of the $\mathcal{N}=4$ SQED with one flavor, regarded as a function of the the background vector multiplet $\hat V$ associated to the  topological symmetry  (in a suitable gauge fixing), is given by
$$
Z_{SQED}[\hat V]=\int DV ~DQ ~e^{\frac{iS_V(V)}{g^2}+iS_{BF}(V,\hat V)+iS_H(V,Q) }\,.
$$
We refer the reader to  \cite{Kapustin:1999ha} for the definition of the kinetic terms for the  vector and hypermultiplet  $S_V(V)$, $S_H(V,Q)$
and record here  the contribution of the BF coupling to the action involving the vector  and the adjoint scalar multiplets $V$,  $\Phi$  and the background linear 
and scalar multiplets $\hat \Sigma$,  $\hat \Phi$:
\be
S_{BF}(V,\hat V)=\int d^3x d^2\theta d^2\bar \theta ~ \hat V \Sigma-\Big(\int d^3x d^2\theta ~i \Phi \hat \Phi+cc\Big)\,.
\ee
The partition function of the free hypermultiplet as a functional of the background vector multiplet  $\hat V$ associated to the  $U(1)$
global symmetry is
$$Z_{H}[\hat V]=\int D Q ~e^{iS_H(\hat V,Q) }\,.$$
In the IR the statement of abelian  $\mathcal{N}=4$ mirror symmetry becomes:
$$\!\!\!\!\!\!\! Z_{SQED}[\hat V]=\int DV ~e^{iS_{BF}(V,\hat V)} ~\int DQ~ e^{iS_H(V,Q) }=
\int D Q~ e^{S_H(\hat V, Q) }=Z_{H}[\hat V]\,,$$
since the  BF coupling action is quadratic it  can be regarded as the  Functional Fourier Transform kernel  and one can  interpret this as the statement that the free hyper  coincides with its Functional Fourier Transform:
$$\int DV ~e^{S_{BF}(V,\hat V)} ~Z_{H}[V]=Z_{H}[\hat V]\,.$$
By using Functional Fourier Transform and the convolution theorem  one can piecewise-generate all the $\mathcal{N}=4$ mirror pairs.

More recently thanks to localisation  this approach has been revamped  in a  simpler although less general fashion.
Partition functions of  $\mathcal{N}=4$ theories on $S^3$ can be computed  via localisation  \cite{Kapustin:2009kz} which reduces them to  matrix integrals, functions of real masses associated to the background vector multiplets of the form $\hat V \sim m \theta \bar \theta$. 
The 1-loop exact  contribution of an hyper to the  $S^3$ partition function is:
$$Z_{H}(\sigma)=\frac{1}{\cosh(\sigma)}\,,$$
and the the basic abelian $\mathcal{N}=4$ mirror symmetry duality is expressed by the integral identity 
$$Z_{SQED}(\xi)=\int d \sigma~ e^{-2\pi i \sigma \xi} \frac{1}{\cosh(\sigma)} = \frac{1}{\cosh(\xi) }  =Z_{H}(\xi) \,.$$
The functional integration becomes an ordinary integration over the locus fixed by the localising equation, that is an integration over the zero modes of the adjoint scalar  $\sigma $ taking value in the Cartan generators of the gauge group. The BF coupling  becomes the ordinary Fourier transform kernel and the basic abelian mirror symmetry reduces to an  ordinary rather than a functional Fourier transform.
Using this trick one can prove all the $\mathcal{N}=4$ mirror dualities.

In the $\mathcal{N}=2$ case the basic Abelian mirror pair is formed by the  $U(1)$ theory with one flavor with $\CW=0$ and the $XYZ$ model:
\be \label{MS2} U(1)_{q,\qt}, \CW=0 \leftrightarrow  \{ x, y, z \}, \CW= x y z \,.\ee
 The chiral ring on the $U(1)$ side is generated by the meson operator $M$ and by the two monopole operators with topological charge $\pm 1$ that we denote $\M^{\pm 1}$ or simply $\M^{\pm}$. 
This duality can also be formulated in a form  equivalent to the $\mathcal{N}=4$ case as 
\be \label{MS4} U(1)_{q,\qt}, \CW=q\qt\Phi \leftrightarrow  \{ p, \pt \}, \CW=0. \ee

In \cite{Dimofte:2011ju} it was observed that  this duality is  the composition of two operations: an  $S$-action which gauges a global  symmetry and introduces a  new topological symmetry and the introduction of a superpotential coupling.

This latter operation has been later named {\it Flip}. If  a theory has a chiral operator $\mathcal O$ with a coupling $\CW=\phi \mathcal O $ to a background chiral $\phi$,
the Flip operation adds a coupling to  a new background chiral field $\phi'$:  $\CW=\phi \mathcal O \to \CW=\phi \mathcal O +\phi \phi' $ and makes $\phi$ a dynamical field.

It is also possible to localise $\mathcal{N}=2$ theories with non trivial R-charges  on the squashed three-sphere $S^3_b$ \cite{Hama:2010av}\cite{Jafferis:2010un}.
The contribution to the partition function of a chiral multiplet  with a real mass $m'$ for a $U(1)$ symmetry and R-charge $r$ is
\be
s_b(\tfrac{iQ}{2}(1-r)-m')\quad {\rm with}\quad s_b(x) = \prod_{m,n {\geq0}}\frac{mb+nb^{-1}+\tfrac{Q}{2}-ix}{mb+nb^{-1}+\tfrac{Q}{2}+ix}\,  ,~\qquad Q=b+1/b\, ,
\ee
where $b$ is the squashing parameter.  For convenience we also introduce the function:
\be
F_m(x)\equiv s_b(x+\tfrac{m}{2} +i\tfrac{Q}{4} )  s_b(-x+\tfrac{m}{2} +i\tfrac{Q}{4} )\ee
that is the contribution of an hyper with  {\it axial}  mass $m$ (using holomorphy \cite{Jafferis:2010un} we absorbed $r$ in a redefinition of the mass).

The fundamental abelian mirror symmetry duality (\ref{MS2}) at the level of the partition  function becomes:
\be
\label{fund}
\int d s e^{-2\pi i s p} F_m(s)=s_b(m) F_{-m}( p )\,.
\ee
Notice that on the r.h.s. the real masses of the three chirals are consistent with the cubic superpotential, the arguments of the three double sine functions sum up to $\tfrac{iQ}{2}$. 

The  identity (\ref{fund}) is known as {\it pentagon identity} and plays a key role in the so called the 3d-3d  correspondence \cite{Dimofte:2011ju},\cite{Dimofte:2011py}. This correspondence relates $3d$ $\mathcal{N}=2$
theories, obtained from the compactification of M5 branes on hyperbolic 3-manifolds $M$, to Chern-Simons theories at complex coupling on $M$. The invariance of  Chern-Simons partition functions under  changes of triangulations corresponds to 3d mirror symmetry and in particular the pentagon identity  represents  the  the basic 2-3 move.\\

By a real mass deformation corresponding to integrating out one chiral, we can derive the mirror duality  between the  $U(1)$ theory with a charge plus  chiral  and $k =+\tfrac{1}{2}$ CS coupling  and a free chiral
with  $k =-\tfrac{1}{2}$ CS coupling. 
In this case one of the two basic monopole operators acquires a non zero gauge charge while the gauge invariant monopole is mapped to the free chiral.
At the level of partition function this duality is encoded in the identity 
\be
\label{csh}
 \int dw ~  e^{\tfrac{-i \pi}{2} w^2}   e^{2\pi i w(  \tfrac{iQ}{4}-p)} s_b(   \tfrac{iQ}{2}-w)=
 e^{\tfrac{i \pi}{2} \left(\tfrac{iQ}{2}-p\right)^2   } s_b(   \tfrac{iQ}{2}-p)\,.
\ee

\subsection*{An example of  mirror duality via partial mirror dualisations}
We will now review  how  to derive the  mirror dual of the  $U(1)$ theory with $K$ flavors $p_i, \pt_i$, $i=1, \ldots, K $ and $\CW=0$, by repeated use of the fundamental Abelian duality via the   Kapustin-Strassler piecewise approach  \cite{Kapustin:1999ha}.

 We dualize each flavor $p_i, \pt_i$, using eq. (\ref{MS4}) from right to left, into a
  $U(1)_{q_i,\qt_i}, \CW=q_i\qt_i\Phi_i$ theory. We end up with a $U(1)^{K+1}$ gauge theory with $2K$ charged  
 chiral fields $q_i,\qt_i$, $K$ gauge-singlets $\Phi_i$ and  a superpotential  $\CW=\sum_i q_i\qt_i\Phi_i$. There are also $2K$ BF couplings  (Chern-Simons interactions involving different $U(1)$ gauge factors) connecting the original $U(1)$ factor with the $2K$ new gauge factors.

\begin{figure}[h]
\centering
\includegraphics[width=3.9in]{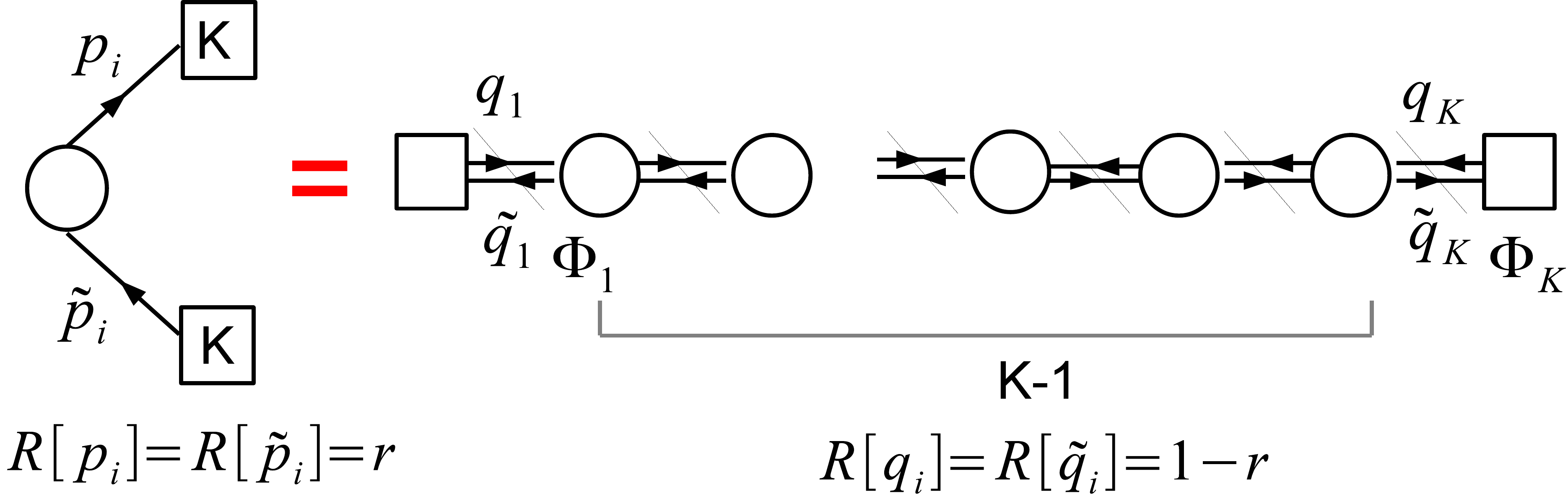}
\caption{The SQED and its quiver dual. On the quiver side in the superpotential  $\CW=\sum_i q_i\qt_i\Phi_i $  each singlet couples to, or flips, the  mesonic operator quadratic in the bifundamental field on its left.  We represent each flip by  slashing the pair of bifundamental fields.  Empty circles (squares) indicate a gauge (flavor) $U(1)$ symmetry.}
\label{sqedquiver}
\end{figure}

At this point the original $U(1)$ gauge group has no flavors and we can perform the functional integral over it obtaining a functional delta function. Implementing the delta reduces the number of gauge groups to $K-1$ and the mirror duality can be presented as
the quiver depicted in  fig. \ref{sqedquiver}. \be \label{MSK} U(1)_{p_i,\pt_i}, \CW=0 \leftrightarrow  [1]-(1)-(1)-\ldots-(1)-[1], \CW=\sum_i q_i\qt_i\Phi_i \ee

\subsection*{Check of the duality at the level of partition function}

We will now retrace the previous  piecewise dualisation at the level of the $S^3_b$ partition function.
For the SQED we turn on real masses in the Cartan generators of 
the global symmetry group  $SU(K)_{m_a}\times SU(K)_{\phi_a} \times U(1)_\Phi\times U(1)_{top, \eta}$
where 
\be
\sum_{a=1}^K m_a=0\,, \qquad  \sum_{a=1}^K \phi_a=\Phi\,.
\ee
The abelian symmetry $U(1)_\Phi$ can mix with the R-charge. The partition function reads:
\be
Z_{SQED}=\int dz e^{2\pi i \eta z} \prod_{a=1}^K F_{-\phi_a} (z+m_a)\,.
\ee
We  now dualize the $K$ hypers by using the  fundamental duality (\ref{fund})  and we integrate over the original $U(1)$
gauge group which has no flavors and yields a delta function. After changing variables
\be
w_1=x_1\,, \quad w_2=x_1+x_2\,, \quad \ldots,  \quad w_{K-1}=\sum_{a}^{K-1}x_a\,,
\ee
we find:
\ben
\label{linmir}
&=&\prod_{a=1}^K s_b(-\phi_a) \int  \prod_{a=1}^{K-1}  dw_a  ~
e^{2\pi i  w_1(m_2-m_1)}e^{2\pi i  w_2 (m_3-m_2)} \cdots e^{2\pi i  w_{K-1} (m_K-m_{K-1})}
e^{-2\pi i  m_K \eta}\nn\\ \nn
&& \times  F_{\phi_1} (w_1)
F_{\phi_2} (w_2-w_1) \cdots F_{\phi_{K-1}} (w_{K-1}-w_{K-2}) 
F_{\phi_K}(\eta-w_{K-1})=  \\
&\equiv&\prod_{a=1}^K s_b(-
\phi_a) \int  \prod_{a=1}^{K-1}  dw_a ~
T_K(\vec  w, \vec m,{\vec\phi},\eta)=Z_{quiver}
 \,.
\een
In this expression the $K$ factors  $s_b(-\phi_a)$ are the contributions of the $K$ chirals singlets  
flipping the bifundamental  mesons.  We also  see that  the real masses are compatible with the 
superpotential ${\mathcal W}=\sum_a^K \Phi_a q_a\tilde q_a$ coupling.

We remark that the check of the duality at  the level of partition functions is for a specific choice of  background hence less general than the Functional Fourier Transform approach. However the partition function approach is very convenient since it allows us to keep track of all  global symmetries across dualities. For example we see that the real masses in the Cartan generators of the flavor $SU(K)_{m_a}$ are mapped  into FI parameters on the mirror quiver side, from which we conclude that the topological symmetry on the quiver theory will enhance.

\subsection*{Mapping of the chiral rings}
We conclude the discussion  of this example by describing the map of the chiral ring generators.
We begin by recording  the formula that computes the global and R-charges of BPS monopole operators (the scaling dimension of the monopole operators coincides with their R-charge). The charge of the monopole operators under 
a global symmetry can be computed by summing over the charges of all the fermions in representations $R_i$ charged under the gauge symmetries \cite{Benna:2009xd}, \cite{Gaiotto:2008ak}, \cite{Bashkirov:2010kz}, \cite{Borokhov:2002ib}, \cite{Borokhov:2002cg}:
\be Q[\M_m]=-\frac{1}{2}\sum_{\psi_i} \sum_{\rho_i \in \mathcal{R}_i} Q[\psi_i]  |\rho_i(m)| \label{Rmon} \,,\ee
where $\rho$ are the weights of the representation.
Let us write this formula more explicitly in the special cases of interest for this paper. In a  quiver  $U(1)^k$ with bifundamentals $B_{ij}$ going from node $i$ to node $j$ and at the $i$-the node  $F_i$  chiral flavors $q_{i,m}$ with gauge charges $\pm 1$, the formula above becomes
\be \label{MONDelta} Q[\M^{(n_1, n_2, \ldots, n_k)}]=
-\frac{1}{2}\sum_{B_{ij}}   Q[B_{ij}]| n_{i}-n_{j}| -\frac{1}{2}\sum_{i=1}^{k}\sum_{m=1}^{F_i}Q[q_{i,m}]|n_{i}|\,, \ee
where  $\M^{(n_1, n_2, \ldots, n_k)}$ is a monopole operator with topological charges $n_1, n_2, \ldots, n_k$.

In the theory $U(1)$ with $N_f$ flavors, the chiral ring is generated by the $N_f^2$ mesons $Q_i \tilde Q_j$ (with $\Delta=2r$, where $r$ is the R-charge and the scaling dimension $\Delta$ of the $N_f$ flavors chiral fields) and by the two magnetic monopoles with topological charges $\pm 1$, $\M^{\pm}$. From eq. (\ref{MONDelta}) we compute $\Delta(\M^{\pm})=-\frac{1}{2}(N_f+N_f)(r-1)=N_f(1-r)$.


In the dual theory $[1]-(1)-(1)-\ldots-(1)-[1]$ the chiral ring contains $N_f$ gauge-singlets $\Phi_i$ with $\Delta=2r$, and  many monopoles operators. We denote the chiral monopoles operators $\M^{(j_1,\cdots, j_{N_f-1})}$, where the $N_f-1$ integers are the charges of topological symmetry of the $i^{th}$ node. There are $N_f(N_f-1)$ special, basic, monopole operators, which generate all the others monopole operators in the chiral ring, they have the lowest possible scaling dimension: $\Delta=2r$. These are the monopole operators with topological charge given by  strings of contiguous $+1$ or $-1$. For example, in the  $4$ flavors case, they can be organized in a matrix
\be \label{monmatrix}
 \left(
 \begin{array}{c c c c }
  & \M^{(1,0,0)} &  \M^{(1,1,0)}& \M^{(1,1,1)} \\
\M^{(-1,0,0)}&   & \M^{(0,1,0)} & \ \M^{(0,1,1)} \\
\M^{(-1,-1,0)}& \M^{(0,-1,0)}&    &   \M^{(0,0,1)} \\
\M^{(-1,-1,-1)}&  \M^{(0,-1,-1)} &  \M^{(0,0,-1)}&   \\
\end{array} 
\right)   \ee
These $N_f(N_f-1)$ operators combine with the $N_f$ flipping singlets  $\Phi_i$ to provide the bifundamental representation of the $SU(N_f) \times SU(N_f)$ enhanced symmetry and are mapped to the $N_f^2$ mesons.
The two remaining generators of the chiral ring are the two long operators
$q_1\cdots q_{N_f}$,    $\tilde q_1\cdots \tilde q_{N_f}$ with $\Delta=N_f(1-r)$ (recall that the bifundamental fields $q_i$ have $\Delta=1-r$) which are mapped to the two monopoles $\M^\pm.$
The generators of the chiral rings perfectly match. 
For example in the $N_f=4$ case we have at level $\Delta=2r$ we have:
\be \label{basicmap}
 \left(
 \begin{array}{c c c c }
p_1 \tilde p_1 & p_1 \tilde p_2 & p_1 \tilde p_3&  p_1 \tilde p_4 \\
p_2 \tilde p_1 & p_2 \tilde p_2 & p_2 \tilde p_3&  p_2 \tilde p_4 \\
p_3 \tilde p_1 & p_3 \tilde p_2 & p_3 \tilde p_3&  p_3 \tilde p_4 \\
p_4 \tilde p_1 & p_4 \tilde p_2 & p_4 \tilde p_3&  p_4 \tilde p_4 \\
\end{array} 
\right)     \leftrightarrow
 \left(
 \begin{array}{c c c c }
\Phi_1 & \M^{(1,0,0)} &  \M^{(1,1,0)}& \M^{(1,1,1)} \\
\M^{(-1,0,0)}& \Phi_2 & \M^{(0,1,0)} & \ \M^{(0,1,1)} \\
\M^{(-1,-1,0)}& \M^{(0,-1,0)}& \Phi_3  &   \M^{(0,0,1)} \\
\M^{(-1,-1,-1)}&  \M^{(0,-1,-1)} &  \M^{(0,0,-1)}& \Phi_4  \\
\end{array} 
\right)      
 \ee
then we have at level $N_f(1-r)$
$$
\M^+ \leftrightarrow \prod_{i=1}^{N_f} q_i \,, \qquad \M^- \leftrightarrow \prod_{i=1}^{N_f} \qt_i\,.
$$

These generators satisfy relations of three types: 
\begin{itemize}
\item[-] The $N_f \times N_f$ meson matrix $p_i\pt_j$ has rank $1$, so every $2 \times 2$ minor of this matrix is vanishing. These relations are simple on the $U(1)$ side, but are non trivial quantum relations on the mirror side. For example in the  $N_f=4$ case we have:
 \ben  \nn\Phi_1 \Phi_2 &=& \M^{(1,0,0)}\M^{(-1,0,0)} \\ \Phi_1\M^{(0,-1,0)}&=&\M^{(1,0,0)}\M^{(-1,-1,0)} \,.\een
\item[-] $\M^+\M^-=0$ in the $U(1)$ theory on  the mirror side becomes: $\prod_{i=1}^{N_f} q_i \prod_{i=1}^{N_f} \qt_i =0$ and follows from  the $\CF$-terms $q_i \qt_i=0$.
\item[-] The relation $\M^{\pm} p_i \pt_j = 0$ is simple to see on the mirror side, for example  if $i=j$: $\prod_{i=1}^{N_f} q_i \cdot \Phi_i = 0$ because of the $\CF$-terms $q_i \Phi_i = 0$.
\end{itemize}

\section{D3 branes ending on the $(1_{NS5} ,  1_{D5'})$ pq-web: a triality}\label{triality}
In this section we study the low energy gauge theory description of the theory living  on two $D3$ branes ending a $(1_{NS5} ,  1_{D5'})$ pq-web, one on each side.
In general by $(H_{NS5},K_{D5'} )$-webs we mean $K_{D5'}$ and $H_{NS5}$  intersecting at a point.
Sometimes  it is convenient to represent the 
$(H_{NS5},K_{D5'} )$-web  by its toric diagram a rectangle with base $K$ and height $H$\cite{Aharony:1997bh}.

Let us briefly review the brane setups. The various branes extend in the directions labelled by $\rm{ x}$ in the table below.
\begin{center}  \begin{tabular}{|c|ccc|ccccccc|}   \hline
& 0 & 1&2&3&4&5&6&7&8&9 \\ \hline
D3 & x&x &x & & & &x & & & \\ 
D5 & x&x &x &x &x &x & & & & \\ 
NS5 & x&x &x & & & & &x &x &x \\ 
D5' & x&x &x &x & & &  & &x &x \\ 
NS5' & x&x &x & &x &x & &x & & \\ 
\hline
\end{tabular}
 \end{center}

The brane setup enjoys a $U(1) \otimes U(1)$ global symmetry that rotates the $45$ and $89$ planes. One combination of these symmetries becomes the $U(1)_R$ global symmetry of the 3d $\CN=2$  gauge theory, while the other combination becomes another   global symmetry that we call $U(1)_t$. We then have the global symmetries inherited from the pq-web. On a $(H_{NS5},K_{D5'} )$-web  the 5d gauge theory is $SU(K)^{H-1}$ linear quiver\footnote{To be more precise, the theory is strongly interacting $CFT_5$ that can be relevantly deformed the $SU(K)^{H-1}$ quiver. Also, using type IIB S-duality, we can deform the $CFT_5$ to the $SU(H)^{K-1}$ quiver.}, and its global symmetries are $SU(K)^2 \otimes SU(H)^2 \times U(1)$.

We choose to terminate the two D3 branes on two NS5', so the brane set up we are interested in is 
\be1_{NS5'} - 1_{D3} - (1_{NS5} ,  1_{D5'}) - 1_{D3} - 1_{NS5'}\ee

The global symmetry of the gauge theory must be 
\be U(1)^4 = U(1)_{NS5'} \otimes U(1)_{(1_{NS5} ,  1_{D5'})} \otimes U(1)_{NS5'} \otimes  U(1)_t \ee

\begin{figure}[ht!]
\centering
\includegraphics[width=3.2in]{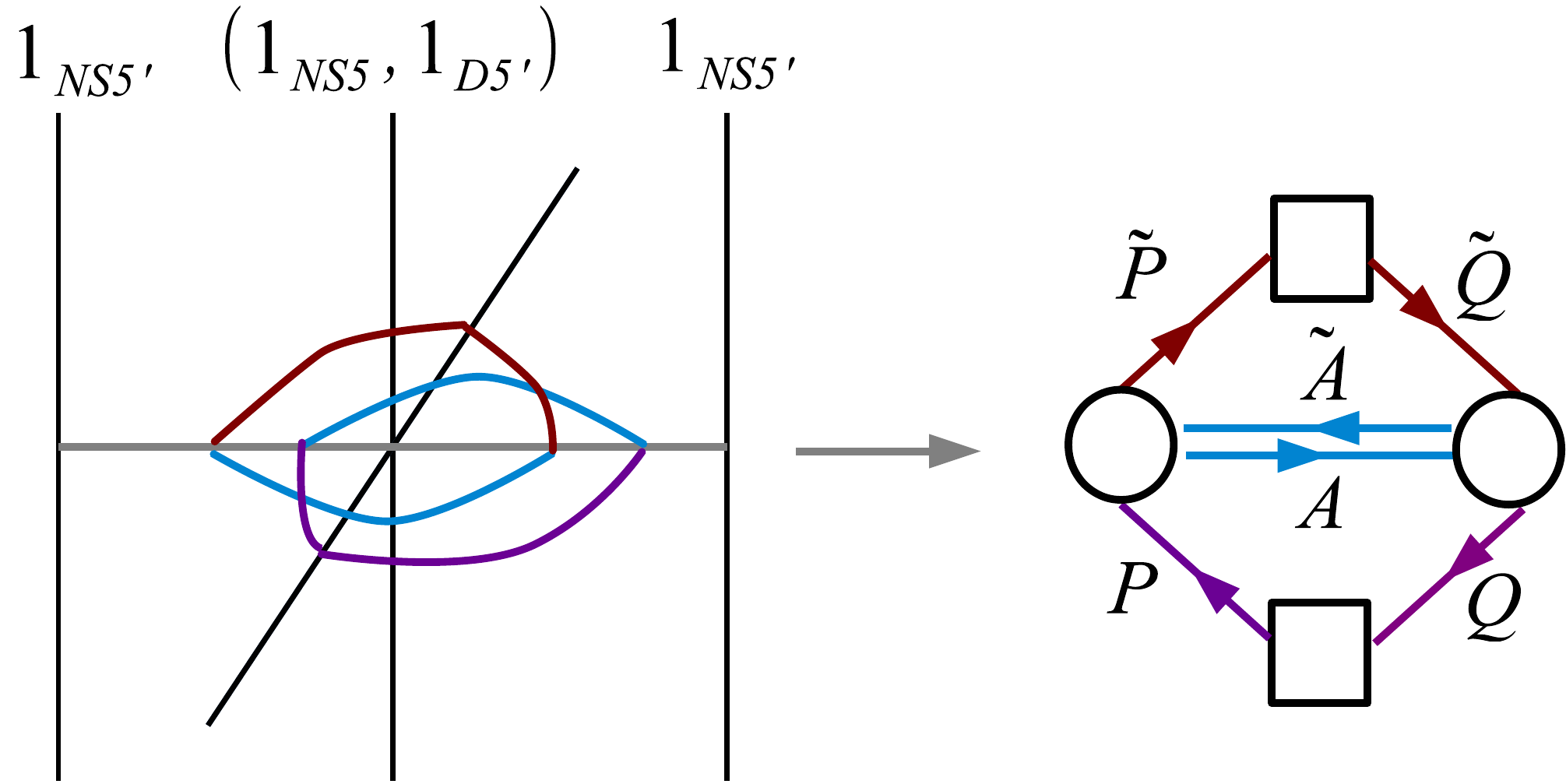}
\caption{The $1_{NS5'} - 1_{D3} - (1_{NS5} ,  1_{D5'}) - 1_{D3} - 1_{NS5'}$ set-up and its low energy quiver gauge theory description. The bifundamental fields  $A,\At$ come from $D3-D3$ strings, the fundamentals  $Q,\Qt$ and $P,\Pt$ come  respectively from $D3-D5'$ strings  and  from $D5'-D3$ strings.
  }
\label{bset}
\end{figure}

The 3d gauge theory description of this brane setup, which we call $\CT_A$ and show in Fig. \ref{bset}, is a $U(1)^2$ quiver theory with 6 chiral fields $(A,\At, Q, \Qt, P, \Pt)$ and two cubic superpotential terms  \cite{Brunner:1998jr}:
\be \CW_{\CT_A} = A P Q + \At \Pt \Qt\,. \ee

There are $6$ fields, so $6$ possible $U(1)$'s, but $2$ combinations are gauged and $2$ combinations are broken by the $2$ superpotential terms, then we have $2$ topological symmetries associated to the $2$ gauge groups, in total we have $U(1)^4$ global symmetry.

The R-charges of the $6$ chiral fields is exactly $2/3$ \footnote{We can see this fact as follows. An alternative presentation of this $U(1)^2$ gauge theory is given by $3$ copies of $U(1)$ with $1$ flavor, supplemented by a functional Dirac delta function that sets to zero the sum of the $3$ gauge fields. In this presentation we can recognize a $\IZ_3$ global symmetry (notice that the superpotential respects this $\IZ_3$), so we can infer that the R-charges of the $6$ fields are all equal. Because of the cubic superpotential, their R-charge is exactly $2/3$.} . Using eq. (\ref{Rmon}) we can compute the scaling dimension of the supersymmetric chiral monopole operators 
\be \Delta[\M_{\CT_A}^{a,b}] = (|a|+|a-b|+|b|)/3\,. \ee

\begin{figure}[h!]
\centering
\includegraphics[width=5.0in]{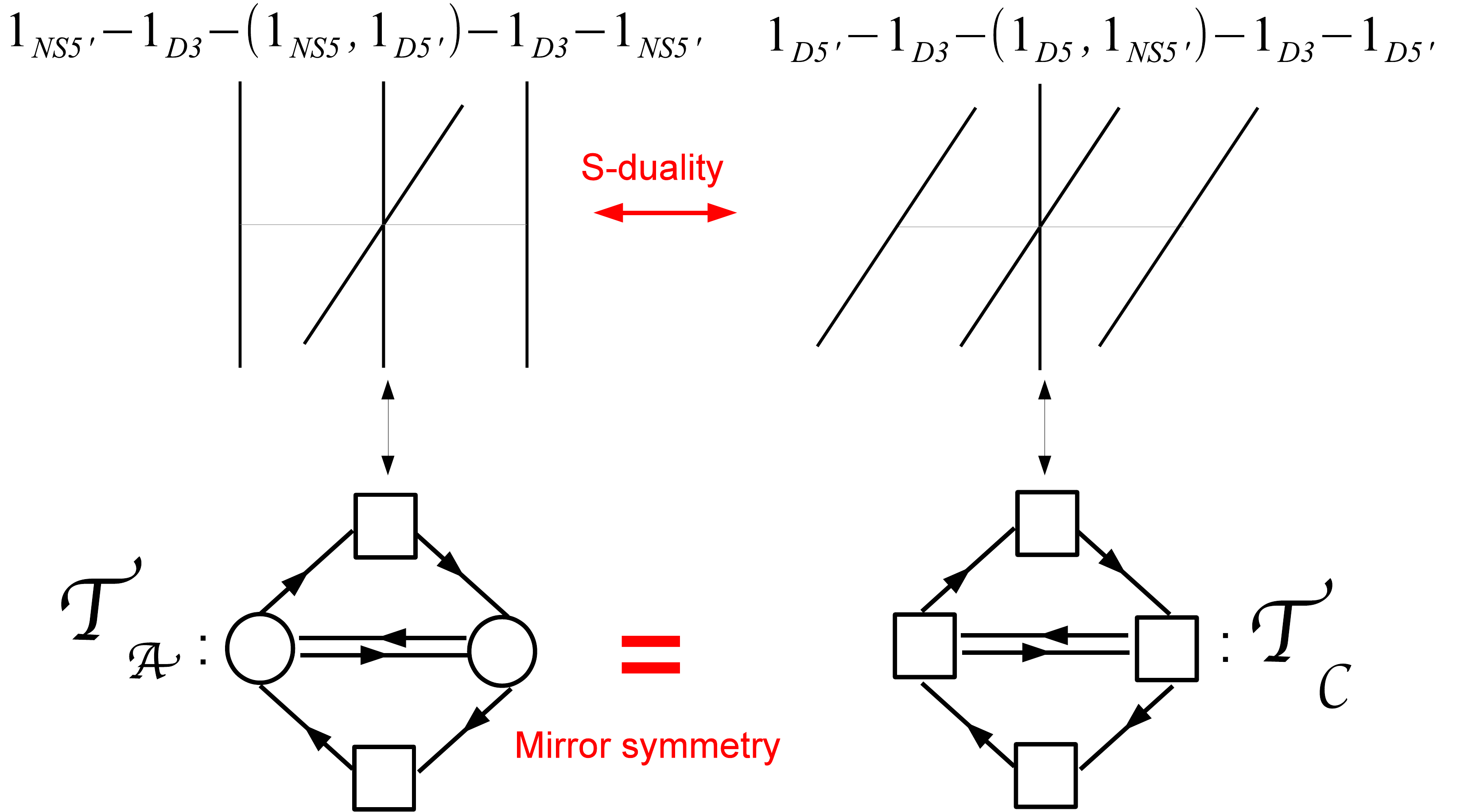}
\caption{
The S-dual brane set-up and the 3d mirror theories. In $\CT_C$ the bifundamental field $a,\at$ come from $D3-D3$ strings, the fundamental $q,\qt$ and $p,\pt$ come respectively from $D3-D5$  and $D5-D3$ strings.}
\label{11web}
\end{figure}

The Type-IIB S-dual brane setup is
\be1_{D5'} - 1_{D3} - (1_{D5} ,  1_{NS5'}) - 1_{D3} - 1_{D5'}\ee
and the 3d description of this brane setup, which we call $\CT_C$ depicted in fig. \ref{11web}, is basically $\CT_A$ without the gauge fields: it's given by $6$ chiral fields, which we denote as $(a, \at, q, \qt, p, \pt)$, entering in two cubic superpotential terms 
\be \CW_{\CT_C} = a p q + \at \pt \qt \,.\ee
Also here the global symmetry is clearly $U(1)^4$, and the scaling dimensions of the $6$ fields is $\Delta=2/3$.

\begin{figure}[h!]
\centering
\includegraphics[width=3.6in]{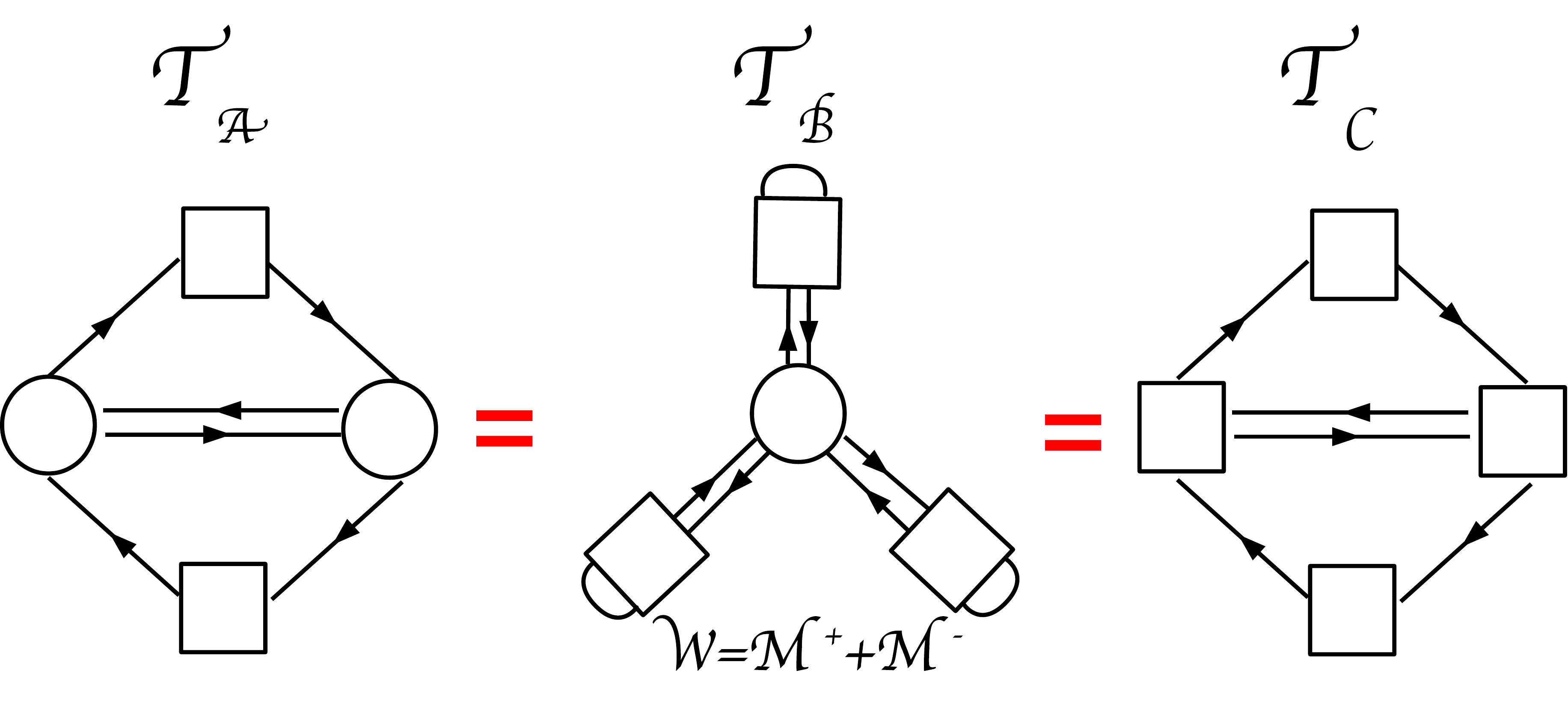}
\caption{The gauge theory triality.}
\label{triweb}
\end{figure}

Our first goal is to show that $\CT_A$ and $\CT_C$ are dual. 
 \footnote{The SQED and its mirror quiver described in the previous section is associated respectively to 
  a $D3$ brane stretching between an $NS5'$ brane and an $(1_{NS5},K_{D5'})$-web and to its S-dual.
There are  some subtleties associated to these  brane setups: they can also correspond to the same gauge theories with additional chiral  singlets entering the superpotential by flipping some or all the mesons (or their mirror image). For instance let us take $K=1$, the system $NS5'-D3-D5'-D3-NS5$ corresponds to $U(1)$ with $1$ flavor. The system $NS5'-D3-NS5-D3-D5'$ corresponds to $U(1)$ with $1$ flavor $q,\qt$, plus a singlet $\Phi$ and $\CW=\Phi q \qt$. These subtleties  are not  present in the  doubled systems which we discuss in this and in the next sections.}
Along the way we will uncover an additional duality with a theory $\CT_B$, an $U(1)$ gauge theory with $3$ flavors and monopole operators in the superpotential, so we will prove the triality shown in Fig. \ref{triweb}.

 \subsection{$\CT_A=\CT_B=\CT_C$: a gauge theory triality}

\subsection*{$\CT_A=\CT_C$}\label{onetwo}

\begin{figure}[h]
\centering
\includegraphics[width=3.2in]{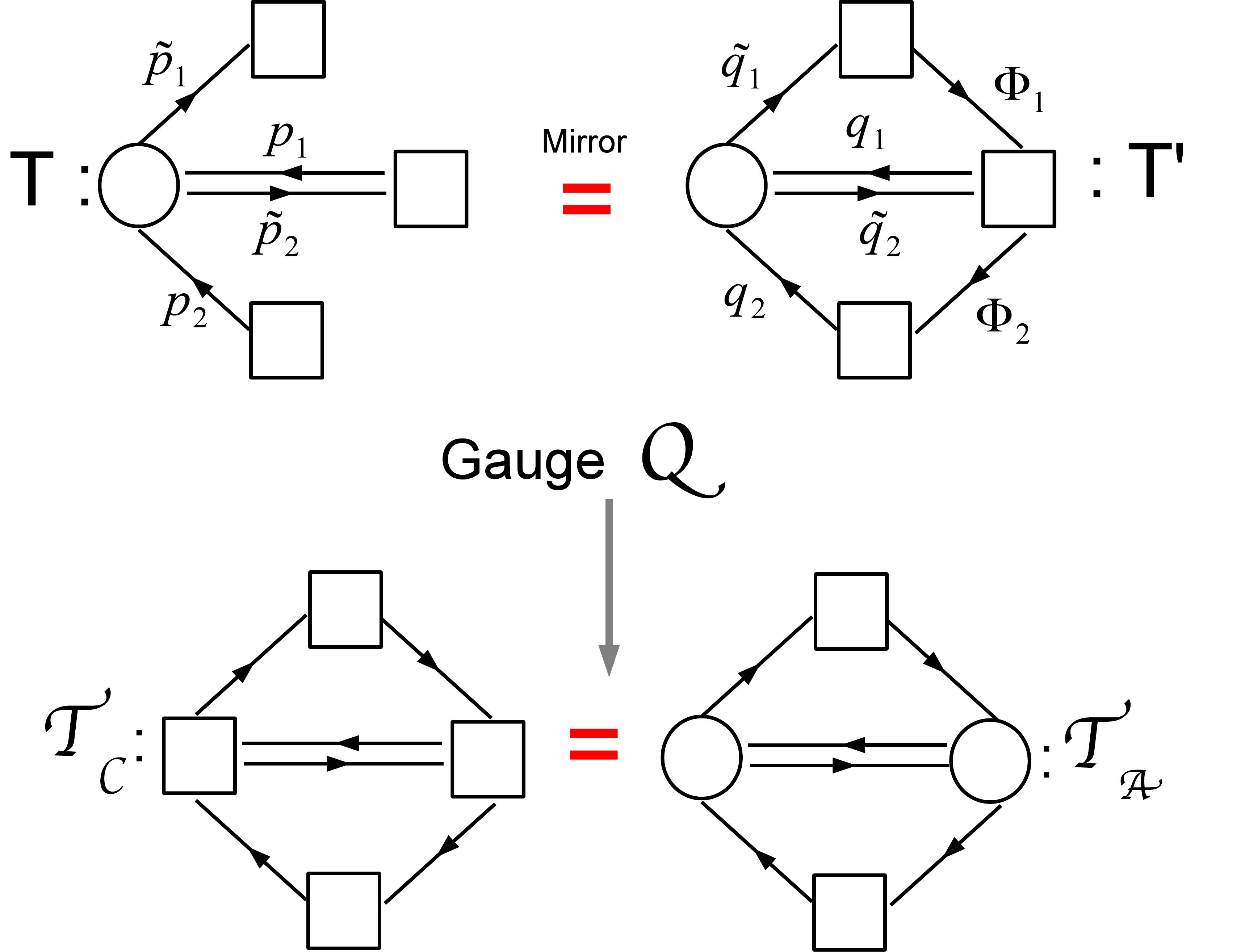}
\caption{Starting from the mirror pair $T=T'$ and gauging the symmetry $\cal {Q}$ yields the $\CT_A=\CT_C$ duality.}
\label{tatc}
\end{figure}

In the quiver $\CT_A$ each $U(1)$ node has  $2$ flavors, so we start with the mirror symmetry for $U(1)$ with 2 flavors reviewed in the previous section and sketched  in the first line of fig.  \ref{tatc}:
 \be T: U(1)_{p_i,\pt_i}, \CW=0 \leftrightarrow T': U(1)_{q_i,\qt_i}, \CW=\Phi_1 q_1 \qt_1 + \Phi_2 q_2 \qt_2\,, \ee 
with the  following mapping of the chiral ring generators:
 \be (p_1 \pt_1, p_2 \pt_2; p_1 \pt_2, p_2 \pt_1; \M^+, \M^-) \leftrightarrow (\Phi_1, \Phi_2; \M^+, \M^-; q_1\qt_2, q_2\qt_1 ) \,.\label{T0T'map}\ee
 The theory $T'$ becomes our theory  $\CT_A$ if we gauge the symmetry $\cal {Q}$ acting on the fields $(q_1, \qt_1, q_2, \qt_2, \Phi_1, \Phi_2)$ with charges $(+1,0,0,-1,-1,+1)$ as shown in fig. \ref{tatc}.
 This implies that the chiral ring generators $(\Phi_1, \Phi_2; \M^+, \M^-; q_1\qt_2, q_2\qt_1 )$ have charges $(-1,+1,0,0,0,0)$ under $\cal {Q}$. 
Using the chiral ring mapping (\ref{T0T'map}), we see that the  symmetry $\cal {Q}$ acts  with charges $(-1,+1,0,0,0,0)$,  on the fields $(p_1\pt_1,p_2\pt_2, p_1\pt_2, p_2\pt_1, \M^+,\M^-)$   of the mirror theory $T$, and in particular  $\cal {Q}$ acts on $(p_1,p_2,\pt_1,\pt_2)$ with charges $(0,1,-1,0)$.\footnote{We could have chosen the charges $(x,x+1,-1-x,-x,0,0)=(0,1,-1,0)+x(1,1,-1,-1)$. The part proportional to $x$ is actually the gauge charge, so it's only a change of basis.}
 
 $\CT_A$ is thus dual to a $U(1)^2$ GLSM with 4 chiral fields $(p_1,p_2,\pt_1,\pt_2)$ and gauge charges vectors $v_1=(1,1,-1,-1)$, $v_2=(0,1,-1,0)$. If we change basis to $w_1=v_1-v_2$, $w_2=v_2$ we can easily see that this GLSM is actually the product of two copies of $U(1)$ with $1$ flavor and $\CW=0$.
 Using that basic mirror symmetry we conclude that $\CT_A$ is mirror of two copies of the $XYZ$ model, that is $\CT_C$.

 In Appendix \ref{LONG} we present a different proof, where the theory remains a quiver at every duality step.

 \subsection*{$\CT_A=\CT_B$}\label{triality1}
\begin{figure}[h]
\centering
\includegraphics[width=4.in]{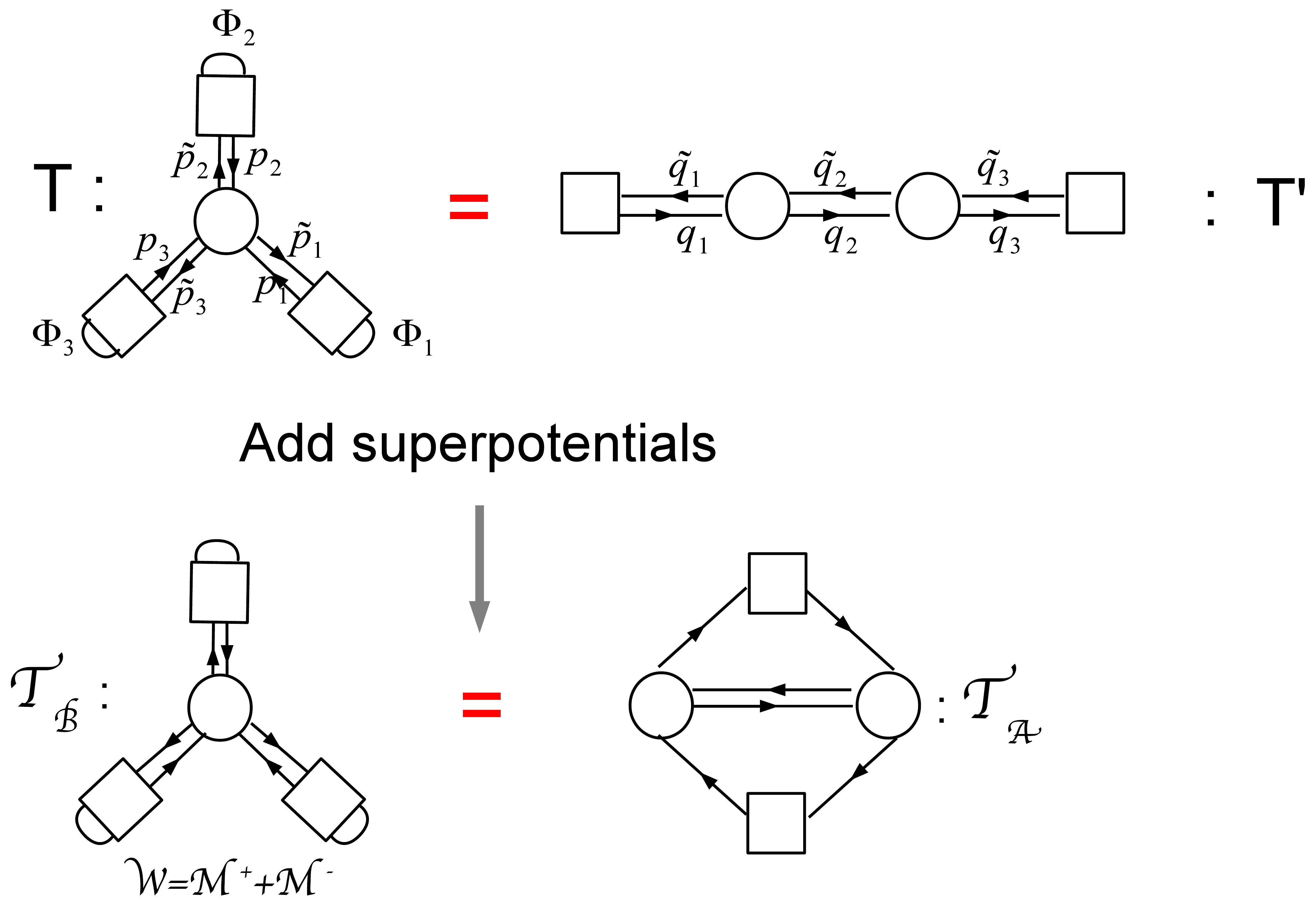}
\caption{Starting from the mirror pair $T=T'$ and adding a cubic superpotential in $T'$ and its dual monopole superpotential in $T$ yields the $\CT_A=\CT_B$ duality. }
\label{tatb}
\end{figure}

Let us start from the known mirror symmetry for $U(1)$ with 3 flavors and flip the 3 diagonal mesons. 
We get a duality between theory $T$, a $U(1)/3, \CW= \sum_i \Phi_i p_i\pt_i$ and theory $T'$ a  $[1]-(1)-(1)-[1]$ linear quiver with $\CW=0$, sketched in the first line of  fig. \ref{tatb}.
 The theory $T'$ is our $\CT_A$ except for the two cubic superpotential terms, which are precisely the dual of the two monopole operators in $T$. So we add the two terms linear in the monopoles to $U(1)/3$, and call the resulting theory $\CT_B$ as shown in fig. \ref{tatb}.
 
We conclude that $\CT_B$, i.e. $U(1)$ with 3 flavors $p_i,\pt_i$ and 
\be \CW =  \sum_i \Phi_i p_i\pt_i + \M^+ + \M^- \,,\ee
is dual to $\CT_A$, the linear quiver $U(1)^2$ with
\be \CW_{\CT_A} = A P Q + \At \Pt \Qt\,. \ee

This proves the gauge theory triality  in fig. \ref{triweb}.

In $\CT_B$ it is easy to compute the R charges: by the $S_3$ symmetry permuting the 3 flavors, it is clear that all the $3$ flavors have the same R charge $r$. Setting the R-charges of the monopoles to be $2$ we get the R-charges of the fundamental fields
\be R( \M^\pm ) = 3(1-r) = 2 \rightarrow r=1/3\ee
In section \ref{rings} we  will discuss the chiral rings of $\CT_A$, $\CT_B$, $\CT_C$ and their mapping.

\subsection*{The theory $U(1)/3, \CW= \M^+ + \M^-$}\label{magicid}
It is interesting to undo the 3 meson flips  in  $\CT_B$ and consider the $SU(3) \times SU(3)$ invariant theory $U(1)/3, \CW= \M^+ + \M^-$. One nice dual phase, with evident $SU(3) \times SU(3)$ symmetry, of this theory is obtained flipping the corresponding fields in $\CT_C$. We obtain a theory of $9$ chiral fields, in order to have $SU(3) \times SU(3)$ symmetry, these $9$ fields must transform in the bifundamental representation, so we call them $x^i_j$, the only cubic superpotential compatible with the symmetry has $6$ terms:
\be \CW = \epsilon_{i j k} x^i_{\tilde 1} x^j_{\tilde 2}x^k_{\tilde 3}\,. \ee
The fields $x^i_j$ have R-charge $2/3$. The duality is
\be \label{magicid} U(1)_{q_i,\qt_i}, \CW=  \M^+ + \M^- \leftrightarrow  \{x^i_j\}, \CW=\epsilon_{i j k} x^i_{\tilde 1} x^j_{\tilde 2}x^k_{\tilde 3} \,.\ee
The mapping of the chiral ring generators is simply
\be q_i \qt_j \leftrightarrow x^i_j \,.\ee
The $F$-terms relations on the right hand side are simply telling us that the $3 \times 3$ meson matrix $q_i\qt_j$ has rank$=1$, a fact that is automatic in the left hand side.

If we now flip back the three diagonal mesons in $U(1)/3$ to get $\CT_B$, on the right hand side of \ref{magicid} we are left with only 6 fields and two superpotential terms, 
\be \CW=  x^2_1 x^3_2 x^1_3 + x^3_1 x^1_2 x^2_3\,.\ee
This is $\CT_C$. The mapping of the chiral ring at level $\Delta=2/3$ reads
  \be   
 \left(   \begin{array}{c c c }
        & p_1 \pt_2 & p_1 \pt_3 \\
p_2 \tilde p_1 &         & p_2 \pt_3 \\
p_3 \tilde p_1 & p_3 \pt_2 &  \\
\end{array}  \right) 
   \leftrightarrow
  \left( \begin{array}{ c c c }
   & x^1_2  &  x^1_3 \\
x^2_1 &  & x^2_3 \\
x^3_1 & x^3_2&   \\
\end{array}  \right)  \,.
  \ee

 \subsection{$S^3$ partition function: the $Z_{\CT_A}=Z_{\CT_B}=Z_{\CT_C}$ identity}

In this section we will derive the triality at the level of the partition function.

\subsubsection*{The $U(1), q_i \tilde q_i$, $\mathcal{W}=\M_++\M_-$  and its mirror, or the new pentagon}

Our starting point is  the   Aharony  duality \cite{Aharony:1997gp} between  theory 
 $\mathcal{T}$, a $U(1), p_i\tilde p_i$  theory with $N_f=2$ and  ${\mathcal W}=0$
 and  theory  $\mathcal{T}'$, a $U(1), q_i\tilde q_i$  theory with $N_f=2$  and ${\mathcal W}=S_{ij} q_i\tilde q_j+
 O^+ \M_++O^- \M_-$.  In $\mathcal{T}'$ the $4$ mesons are flipped by singlets $S_{ij}$ and the monopoles by singlets $O^\pm$.
We can prove this duality starting from the partition function of $\CT$:
\be
Z_{\CT}=\int dx e^{2\pi i x \xi} F_{m_1}(x+\tfrac{\mu}{2}) F_{m_2}(x-\tfrac{\mu}{2})\,,\nn
\ee
and applying 6 times the pentagon identity (\ref{fund}) to get:
\ben
\label{quiver}
\nn Z_{\CT}=&&
\!\!\!\!\!\!\!\!= 
e^{ \tfrac{\pi i\xi(m_2-m_1)}{2}}~
s_b(m_1) s_b(m_2)
 s_b\left(\tfrac{(m_2+m_1)}{2} \pm \mu\right)
 s_b\left(-\tfrac{(m_2+m_1)}{2}  \pm \xi\right)
 \\&& 
\times \int dw ~e^{2\pi i w \xi}~ F_{-\tfrac{(m_2+m_1)}{2}  +\mu}(w) F_{-\tfrac{(m_2+m_1)}{2} - \mu}\left(w+\tfrac{(m_2+m_1)}{2} \right)=Z_{\CT'}\,.
\een

Notice the six singlets flipping the mesons and the monopoles operators.
Now we act  on both sides with  the $S$ element, which  gauges   the $U(1)_\xi$ symmetry and introduces a new FI parameter
$\eta$:
\be
\int d\xi  e^{2\pi i \xi \eta} Z_{\CT}=\int d\xi  e^{2\pi i \xi \eta} Z_{\CT'}\,,
\ee
by rearranging we find:
\ben
\label{tbtc}
&&Z_{\CT_B}=\nn
s_b(\tfrac{(m_2+m_1)}{2}  \pm \mu)   s_b(-(m_1+m_2)-\tfrac{iQ}{2})  \times \\
&&\nn \int dw  ~ F_{(m_1+m_2)+\tfrac{iQ}{2}} (w+\eta)
  F_{-\tfrac{(m_2+m_1)}{2}+\mu }\left(w -\tfrac{(m_2-m_1)}{4}\right) F_{-\tfrac{(m_2+m_1)}{2}  - \mu}\left(w+\tfrac{(m_2-m_1)}{4}\right)=
\\ &&=  s_b(-m_1) s_b(-m_2) F_{m_1}(\tfrac{\mu}{2}-\eta)F_{m_2}(\tfrac{\mu}{2}+\eta)     =Z_{\CT_C}
     \,.
 \een
We have identified the l.h.s. and r.h.s. of this identity with the partition functions of theories $\CT_B$ and $\CT_C$. 
Indeed $Z_{\mathcal{T}_B}$ has no real mass for the topological  symmetry  and the masses for the flavor symmetry satisfy a constraint.
This is compatible with the  breaking of $U(1)_{top}\times U(1)_{axial}$ by the a superpotential linear in the monopole operators $\CW=  \M_++\M_-$.
The three flipping singlets are also visible.
In $Z_{\mathcal{T}_C}$ we have  6 chirals with real masses compatible with two cubic superpotentials.

If we now introduce the following parameterization 
\ben
\nn&&a_1,b_1=-iQ/4+\tfrac{\mu}{2}-\tfrac{(m_1+m_2)}{4}\mp  \tfrac{(m_2-m_1)}{4} \\
\nn&&a_2,b_2=-iQ/4-\tfrac{\mu}{2}-\tfrac{(m_1+m_2)}{4}\pm \tfrac{(m_2-m_1)}{4}\\
&&a_3,b_3=\tfrac{(m_1+m_2)}{2}\pm \eta\,,
\een
 we can rewrite eq. (\ref{tbtc}) as the following  identity
\be
\label{newpent}
 \int ds  \prod_{i=1}^3 s_b(\tfrac{iQ}{2}+a_i+s)  s_b(\tfrac{iQ}{2}+b_i-s)=\prod_{i,j=1}^3s_b(\tfrac{iQ}{2}+a_i+b_j)
\ee
with
\be
\sum_{i=1}^3(a_i+b_i)=-i Q\,.
\label{conpent}
\ee
On the l.h.s. we can shift the integration variable sending $a_i \rightarrow a_i + \Delta$ and $b_i \rightarrow b_i - \Delta$. Similarly the r.h.s.  depends  only on the combination $a_i+b_j$ invariant under the previous shift. By considering also the constraint (\ref{conpent}) we see that indeed the  identity depends only on $4$ parameters, which are associated to the Cartan generators of the $SU(3)^2$ global symmetry.
Notice also that on the l.h.s. of eq. (\ref{newpent})  there is no FI parameter, since the topological symmetry is broken by the monopole superpotential.  This is the partition function identity  for the  duality  we found in (\ref{magicid}).

This integral identity appears in various contexts in the  mathematical literature, for example when derived in \cite{Volkov2003} in  was called {\it ultimate integral identity}. More recently it has been called {\it new pentagon identity} and
plays a  role in the study of shaped triangulations, \cite{Kashaev:2012cz}. 
This identity can also be obtained as a particular limit of the elliptic beta-integral discovered by Spiridonov \cite{spiri} and has been tested 
also on   the 3d  index  \cite{Gahramanov:2013rda}.

A similar relation for $U(1)$ with $4$ flavors and monopole operators in the superpotential has been discussed in \cite{Dimofte:2012pd} and plays an important role in the 3d-3d correspondence. Both of these dualities can be generalized to a Aharony-Seiberg duality for $U(N_c)$ with $N_f$ flavors, and $\CW= \M^+ + \M^-$ \cite{toappear1}.

\subsubsection*{Partition functions triality}

We now complete the partition function proof of the triality.
We start from theory $\mathcal{T}_A$ where we turn on  real masses for  the global symmetry $U(1)_{top,\xi_1}\times U(1)_{top,\xi_2}\times U(1)_{t}\times U(1)_\phi $. In the table below we record the charges of the fields in fig. \ref{bset} and their contribution to the partition function:
 \begin{center}  \begin{tabular}{|c|c|c|c|c|c|c|c|}   \hline
    & $U(1)_{x}$ & $U(1)_{y}$  & $U(1)_{\phi}$     & $U(1)_t$ &$U(1)_{R_o}$& $U(1)_R=\Delta$ & Z\\ \hline
    $A$ & -1 & 1                            & $0$                 & $1$          &  $1/2$            & $2/3$      & $s_b(\tfrac{iQ}{4}+t+(x-y))$\\ \hline
    $\At$ & 1 & -1                          & $0$                 & $1$          & $1/2$            & $2/3$          & $s_b(\tfrac{iQ}{4}+t-(x-y))$ \\ \hline
    $P$ & 1& 0                             & $-1/2$              & $-1/2$      &$3/4$           & $2/3$           & $s_b(\tfrac{iQ}{8}-\tfrac{t}{2}-\tfrac{i\phi}{2}-x)$\\ \hline
    $\Pt$ & -1 & 0                         & $-1/2$              &$-1/2$        &$3/4$            & $2/3$         &  $s_b(\tfrac{iQ}{8}-\tfrac{t}{2}-\tfrac{i\phi}{2}+x)$\\ \hline
    $Q$ & 0 & -1                           & $1/2$                &$-1/2 $        & $3/4$         & $2/3$          & $s_b(\tfrac{iQ}{8}-\tfrac{t}{2}+\tfrac{i\phi}{2}+y)$ \\ \hline
    $\Qt$ & 0 & 1                        & $1/2$               & $-1/2$         & $3/4$         & $2/3$         & $s_b(\tfrac{iQ}{8}-\tfrac{t}{2}+\tfrac{i\phi}{2}-y)$\\ \hline
      \end{tabular}\label{chargesT1} \end{center}
  The trial R-charges under $U(1)_{R_o}$ are consistent with the superpotential couplings indeed one can verify that    
the  real masses of $P,Q,A$ and $\tilde P, \tilde Q,\tilde A$  sum up to $\tfrac{iQ}{2}$.
Notice that the abelian symmetry $U(1)_t$ mixes with $U(1)_{R_o}$ to give $U(1)_R=U(1)_{R_0}+\tfrac{1}{6} U(1)_t$.
 
Using this table it is easy to write the partition function of theory $\CT_A$:
\ben
\label{a}
 Z_{\CT_A}&=&
  \int dx_1 dx_2 e^{2\pi i (\xi_1x_1+\xi_2 x_2)} 
F_{2t}(x_2-x_1)  F_{2v-\phi}(x_1)  F_{2v+\phi}(x_2)\,,
 \een
 where we combined pairs of chirals of opposite gauged charges into hypers and introduce for convenience the parameter  $2v=-t-iQ/4$.

 Now we dualize each hyper by using the pentagon identity (\ref{fund}):
\ben
\label{a=b}
&=&
 s_b(2t) s_b(2v\pm \phi)  \int dx_1 ~dx_2 ~ds~ dp ~dq ~e^{2\pi i (\xi_1x_1+\xi_2 x_2)} 
e^{-2\pi(x_2-x_1)s} e^{-2\pi x_1 p}e^{-2\pi x_1 q}\nn\\
&&F_{-2t}(s)  F_{-2v+\phi}( p ) F_{-2v-\phi}(q)\,.
 \een
Integrations over $x_1,x_2$ give $\delta(\xi_1-p+s)$ $\delta(\xi_2-q-s)$, which we implement by integrating over $p,q$, in the end we find:
\ben
&=&
s_b(2t) s_b(2v\pm \phi)  \int ds F_{-2t}(s)  F_{-2v+\phi}( s+\xi_1 ) F_{-2v-\phi}(\xi_2-s)=Z_{\CT_B}\,.
 \een
In the last step we  identified  the partition function of theory $Z_{\CT_B}$, the $U(1)$ theory with 3 flipped flavors and a superpotential
 $\mathcal{W}=\M_++\M_-+\sum_i^3  \Phi_i a_i b_{\bar i}$.
This proves the first part of the triality, which follow by piecewise mirror symmetry.
%
%
%
%
%
The triality can be summarised as follows:
  \ben
  \label{ftri}
&&\!\!\!\!\!\!\!\!\!\!\!
Z_{\CT_A}=\int dx_1 dx_2 e^{2\pi i (\xi_1x_1+\xi_2 x_2)} 
F_{2t}(x_2-x_1)  F_{2v-\phi}(x_1)  F_{2v+\phi}(x_2)= \nn\\
&&= s_b(2t) s_b(2v\pm \phi)  \int ds F_{-2t}(s)  F_{-2v+\phi}( s+\xi_1 ) F_{-2v-\phi}(\xi_2-s)=Z_{\CT_B}
\nn\\
&&=F_{2t}(\xi_1+\xi_2)  F_{2v+\phi}(\xi_1)  F_{2v-\phi}(\xi_2)=Z_{\CT_C}\,.
\een
As we have just shown  $Z_{\CT_A}=Z_{\CT_B}$ follows from the pentagon identity (\ref{fund}) while the equality   $Z_{\CT_B}=Z_{\CT_C}$ is due to  the new pentagon identity.
Finally  $Z_{\CT_A}=Z_{\CT_C}$ gives a new interesting  identity:
 \be
 \label{magfor0}
\int dx_1 dx_2 e^{2\pi i (\xi_1x_1+\xi_2 x_2)} 
F_{2t}(x_2-x_1)  F_{2v-\phi}(x_1)  F_{2v+\phi}(x_2)= F_{2t}(\xi_1+\xi_2)  F_{2v+\phi}(\xi_1)  F_{2v-\phi}(\xi_2)\,.
 \ee
We can introduce the function:
\be
B(x_1,x_2,v,\phi ):=F_{2t}(x_1-x_2)  F_{2v-\phi}(x_1)  F_{2v+\phi}(x_2)\,,
\ee
which satisfies:
\be
B(x_1,x_2,v,-\phi )=B(x_2,x_1,v,\phi )\,,
\ee
and
\be
\label{Bide}
\int dx_1 dx_2 e^{2\pi i (\xi_1x_1+\xi_2 x_2)} 
B(x_1,x_2,v,\phi )=B(-\xi_1,\xi_2,v,-\phi )\,.
 \ee
This Fourier-transform-like identity describing the S-duality of the  brane setup  in fig. \ref{11web} (given that the $(1,1)$ pq-web is self dual) is very reminiscent of the S-transform action on codimension-two defects discussed in   \cite{Gaiotto:2014ina}.

\subsection{The full structure of the three chiral rings}\label{rings}
In the previous sections we proved that the three theories $\CT_A, \CT_B, \CT_C$ are dual, using Kapustin-Strassler mirror symmetry arguments, this implies  in particular that the chiral rings of the three theories  must be isomorphic. It is however instructive to study the relations in three chiral rings separately. This section is not necessary for the following and it can be skipped.

The chiral ring of $\CT_C$, which is Wess-Zumino model $apq+\at\pt\qt$ with no gauge interactions, is very simple to compute. It is generated  by the $6$ fields $a,p,q, \tilde a,\tilde p, \tilde q$ with  $\Delta=2/3$.


The  $\CT_A$  chiral ring contains monopole operators  with charges $a,b$ under the topological symmetries $U(1)_{top,\xi_1}\times U(1)_{top,\xi_2} $ and dimension
\be
\Delta[\M_{\CT_A}^{a,b}]=(|a|+|a-b|+|b|)/3\,.
\ee
The  $6$ $\Delta=2/3$ monopole operators 
\be \M^{1,0}, \M^{0,1}, \M^{1,1}, \M^{-1,0},\M^{0,-1},\M^{-1,-1}\,,\ee
dual to the $\CT_C$ chiral ring generators are the $\CT_A$ chiral ring generators. 
As we shall see the  relations in $\CT_A$ involve non trivial quantum relations.

Finally in  $\CT_B$  there are  $6$ non vanishing mesons with $\Delta=2/3$ which generate the chiral ring.

We can use the two $\CT_A$ topological charges to organize the structure of the chiral rings.
For example at level $\Delta=2/3$ the mapping of the  $\CT_A \leftrightarrow \CT_C \leftrightarrow  \CT_B$ chiral ring generators is conveniently written as
   \be   \left(
 \begin{array}{ c c c }
   & \M^{1,0} &  \M^{1,1} \\
\M^{0,-1}&  & \M^{0,1} \\
\M^{-1,-1}& \M^{-1,0}&   \\
\end{array}  \right)      
     \leftrightarrow
     \left(
 \begin{array}{ c c c }
   & p  &  \at \\
\qt &  & q \\
a & \pt&   \\
\end{array}  \right)      
     \leftrightarrow
 \left(   \begin{array}{c c c }
        & p_1 \pt_2 & p_1 \pt_3 \\
p_3 \tilde p_2 &         & p_2 \pt_3 \\
p_3 \tilde p_1 & p_2 \pt_1 &  \\
\end{array}  \right)   \ee

In the next sections we will discuss the relations in the three chiral rings.

\subsection*{$\CT_C$ chiral ring}
In theory $\CT_C$ the 6 generators obey  $6$ quadratic $\CF$-terms relations 
\be \label{6rel} ap=pq=qa= \at\pt=\pt\qt=\qt\at=0 \,,\ee
coming from the cubic superpotential $\CW_{\CT_C} = a p q + \at \pt \qt$. 

 At the second level, $\Delta=4/3$, we can construct $\frac{6\cdot5}{2}=21$ operators, but $6$ of these are vanishing, we are left with $15=12+3$ operators, that can be organized  as follows:
 \be   \label{LEV2chiralring}   \left(
 \begin{array}{ c c c c c}
    &     &  \,\, p^2 \,\, & \,\, p\at \,\, & \,\, \at^2 \,\, \\
    & p\qt&     &      & q\at \\
\qt^2& & a\at,p\pt,q\qt  &  & q^2 \\
a\qt & &  & q\pt &   \\
\,\, a^2 \,\,  & \,\, a\pt \,\, &  \,\, \pt^2 \,\, &  &   \\
\end{array}  \right)     \ee

The operators in the chiral rings can be thought of as the holomorphic functions on the moduli space of vacua, which is simply the product of the moduli space of vacua of two $XYZ$ models:
\be (\mathbb{C}_a \oplus \mathbb{C}_p \oplus \mathbb{C}_q) \otimes  (\mathbb{C}_\at \oplus \mathbb{C}_\pt \oplus \mathbb{C}_\qt)\ee

It's also easy to compute the Hilbert Series \cite{Benvenuti:2006qr} of the chiral ring. The (unrefined) Hilbert Series for a single $XYZ$ model is
\be \HS_{XYZ} = \frac{3}{1-t}  - 2 \ee
where we need to subtract $2$ to avoid overcounting the identity operator. The Hilbert Series for the product of $2$ $XYZ$ models is thus
\be \HS_{\CT_C} = \left( \frac{3}{1-t} - 2 \right)^2 = \frac{1+4t+4t^2}{1-t^2} \ee
which admits the small $t$ expansion
\be \HS_{\CT_C} = 4 + \sum_{n=0}^\infty (9k-3)t^k = 1 + 6t + 15 t^2 + 24 t^3 + \ldots \ee

From the last equation we can infer that at level $k$ there are $6k$ operators living on the edges of an hexagon in the $\mathbb{Z}\times \mathbb{Z}$ lattice with vertices $\pm(k,0),\pm(k,k),\pm(0,k)$, and, for $k>0$, $3k-3$ operators living inside this hexagon, along the $3$ main diagonals of the hexagon.
 
\subsection*{$\CT_B$ chiral ring}
In $\CT_B$ there are $3$ basic relations 
\be p_1\pt_1=p_2\pt_2=p_3\pt_3=0 \,, \ee 
which are $\CF$-terms of the $\sum_i \Phi_i p_i\pt_i$ part of the superpotential. These imply that there are $9$ vanishing quadratic relations obeyed by the $6$ mesons $M_{ij} = p_i\pt_j , i \neq j$. For instance $M_{31} M_{12} = p_3 \pt_1 \cdot p_1 \pt_2 = 0$ as a consequence of $p_1 \pt_1 =0$. These are the $6$ relations (\ref{6rel})  in $\CT_C$, in addition we have  $3$ relations
\be M_{12}M_{21} = M_{13}M_{31} = M_{23}M_{32} = 0 \,.\ee 
Moreover, in $\CT_B$ there are $3$ additional chiral ring generators, the $3$ gauge-singlets $\Phi_i$, that have $\Delta=4/3$. 

The $\CT_B$ chiral ring is generated by $6$ $\Delta=2/3$ operators and by $3$ $\Delta=4/3$ operators. As we have seen the $6$ $\Delta=2/3$ operators  obey $9$ quadratic relations. Considering the $\CF$-term of the $\sum_i \Phi_i p_i\pt_i$ of the form $\Phi_i p_i = \Phi_i \pt_i =0$, we find additional relations at level $\Delta=5/3$: $\Phi_i M_{ij}=0$.
Hence, even if at first it looks like that the $\CT_B$ and  $\CT_C$ chiral rings are different, they are just two different presentation of the same algebraic structure. For instance let us look at the level $\Delta=4/3$ operators, there are $15=12+3$ operators, that can be organized  
as:
 \be      \left(
 \begin{array}{ c c c c c}
    &     &  \,\, M_{12}^2 \,\, & \,\, M_{12}M_{13} \,\, & \,\, M_{13}^2 \,\, \\
    & M_{32}M_{12}&     &      & M_{13}M_{23} \\
M_{32}^2& & \Phi_{i=1,2,3}  &  & M_{23}^2 \\
M_{31}M_{32} & &  & M_{23}M_{21} &   \\
\,\, M_{31}^2 \,\,  & \,\, M_{31}M_{21} \,\, &  \,\, M_{21}^2 \,\, &  &   \\
\end{array}  \right)  \,,   \ee
and immediately matched   to the corresponding generators in $\CT_C$ (\ref{LEV2chiralring}).

\subsection*{$\CT_A$ chiral ring}
The chiral ring structure is more involved to uncover in the case of $\CT_A$. The $6$ quadratic relations must be quantum relations involving monopole operators, it turns out these relations are setting to zero the $6$ monopole operators $\M^{x,y}$ with $|x|+|x-y|+|y|=2$, for instance
 \be\M^{1,1}=\M^{1,0}\M^{0,1}=0\,.\ee
 
There are $3$ additional relations. The 3 gauge invariant mesons which are not vanishing in the chiral ring are quadratic in the monopoles:
 \be (Q\Qt, P\Pt, A\At)=(\M^{1,0}\M^{-1,0},\M^{0,1}\M^{0,-1},\M^{1,1}\M^{-1,-1}) \ee
 We can understand the first relation $Q\Qt=\M^{1,0}\M^{-1,0}$ in the following way. Think of $\CT_A$ as a gauging of $T'$ as in fig. 
 \ref{tatc}, i.e. $U(1)$ with $2$ flavors $A,\At$ and $P,\Pt$ where the mesons $AP$ and $\At \Pt$ are flipped by the fields $Q$ and $\Qt$. The  dual of $T'$ is $T$, i.e. $U(1)$ with $2$ flavors $\{p_i,\pt_i\}$ with $\CW=0$. In $T$ the $4$ mesons $M_{ij}$ satisfy the rank-$1$ condition
 \be M_{11} M_{22} = M_{12} M_{21} \,,\ee
 which in $T'$ reads $ Q \Qt = \M^+ \M^- $.
 After gauging the $U(1)$ symmetry to go back to $\CT_C$, this equation becomes $Q\Qt=\M^{1,0}\M^{-1,0}$.

In $\CT_A$ the  $15$ operators at level $\Delta=4/3$ are the 12 monopole operators $\M^{x,y}$ with $|x|+|x-y|+|y|=4$, which can be constructed like $\M^{2,0}=\M^{1,0}\M^{1,0}$ plus  $3$ operators with vanishing topological charge
 $(\M^{1,0}\M^{-1,0},\M^{0,1}\M^{0,-1},\M^{1,1}\M^{-1,-1})$ which sit at the origin of the lattice. Because of the  $3$ relations discussed above, the $3$ mesonic would-be-generators $(Q\Qt, P\Pt, A\At)$ can be eliminated from the presentation of the chiral ring algebraic structure.

It would be nice to apply the formalism developed in \cite{Cremonesi:2013lqa, Cremonesi:2014kwa, Cremonesi:2014vla, Hanany:2015via,Cremonesi:2015dja} and perform a direct computation of the Hilbert Series.

\subsection*{Comments on the triality}
In this section we gave a proof that the $3$ gauge theories $\CT_A, \CT_C, \CT_B$ are dual. The theory $\CT_A$ is associated to the brane setup $1_{NS5'} - 1_{D3} - (1_{NS5} ,  1_{D5'}) - 1_{D3} - 1_{NS5'}$ while we argued that the  mirror dual theory $\CT_C$ is associated to the S-dual brane setup $1_{D5'} - 1_{D3} - (1_{D5} ,  1_{NS5'}) - 1_{D3} - 1_{D5'}$. A natural question is the relation between theory $\CT_B$ and the brane setups. In the next section  we will generalize this result to the setups $1_{NS5'} - 1_{D3} - (1_{NS5} ,  K_{D5'}) - 1_{D3} - 1_{NS5'}$ and its S-dual $1_{D5'} - 1_{D3} - (1_{D5} ,  K_{NS5'}) - 1_{D3} - 1_{D5'}$. We will find that we can associate to each brane setup two slightly different gauge theories, and the difference consists precisely in replacing a subquiver $\CT_B$ with a subquiver $\CT_C$. So in a sense the theory $\CT_B$ is associated to both brane setups $1_{NS5'} - 1_{D3} - (1_{NS5} ,  1_{D5'}) - 1_{D3} - 1_{NS5'}$ and $1_{D5'} - 1_{D3} - (1_{D5} ,  1_{NS5'}) - 1_{D3} - 1_{D5'}$.

The fact that we can do this replacement  without introducing BF couplings  is non-trivial, since in general if we perform a partial dualization we do get  BF couplings. For example if  in a quiver theory we dualize  a pair of chiral fields to a  $U(1)$ theory with $1$ flavor, we also need to add BF couplings between the new $U(1)$ node and all the nodes under which the $2$ chirals were charged.

\section{D3 branes ending on the $(K_{D5'} ,  1_{NS5}) \leftrightarrow  (1_{D5} ,  K_{NS5'}) $ pq-webs}\label{k1web}
In this section we will extend the  previous results to  the  following brane setup:
\be
\label{bsettak}  1_{NS5'} - 1_{D3} - (K_{D5'} ,  1_{NS5}) - 1_{D3} - 1_{NS5'}\ee
and the  S-dual 
\be 
\label{sbsettak} 
 1_{D5'} - 1_{D3} - (1_{D5} ,  K_{NS5'}) - 1_{D3} - 1_{D5'} \ee
The global symmetry  of the first brane set up is
\be U(1)_{1_{NS5'}} \times [SU(K) \times SU(K) \times U(1)]_{(K_{D5'} ,  1_{NS5})} \times U(1)_{1_{NS5'}} \times U(1)_t\,, \ee
and the low energy description  on the $D3$ branes is the well known quiver gauge theory $\CT_{A, K}$ which we discuss in the next section.
The gauge theory corresponding to the second brane setup is not known. We will find it using partial mirror symmetry, i.e. dualizing all the $4K+2$ chiral fields of $\CT_{A, K}$.

We leave the discussion on the low energy description of $D3$ branes ending on the $(K_{D5'} ,  H_{NS5})$ pq-web for the future.

\subsubsection*{$\CT_{A, K}$: global symmetries and chiral ring}
\begin{figure}[h]
\centering
\includegraphics[width=6.0in]{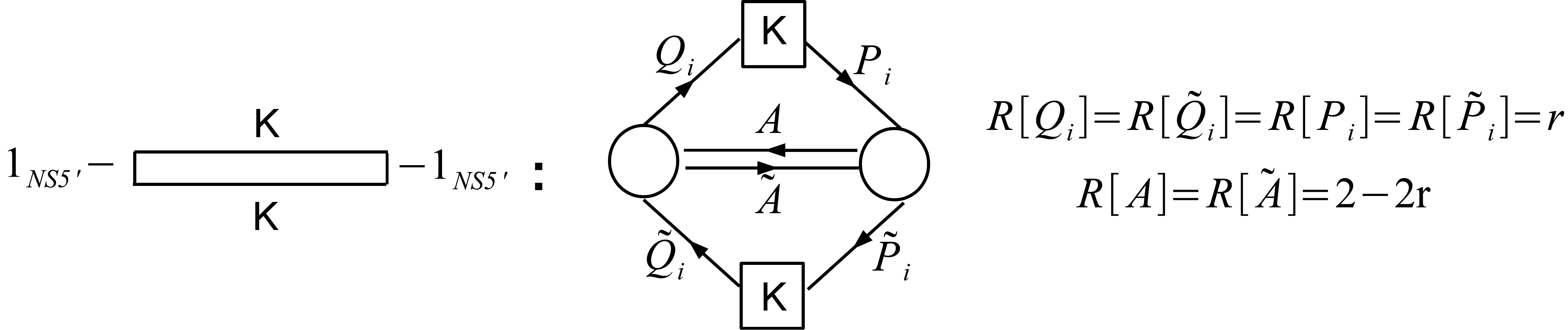}
\caption{The brane set up $1_{NS5'} - 1_{D3} - (K_{D5'} ,  1_{NS5}) - 1_{D3} - 1_{NS5'}$ and its low energy quiver description 
 $\CT_{A, K}$ quiver. The $(K_{D5'} ,  1_{NS5})$-web is depicted by its toric diagram: a rectangle with base $K$ and hight $1$.}
\label{tak}
\end{figure}

We begin with the brane set-up  \ref{bsettak} and study its low energy description $\CT_{A, K}$. This the $U(1) \times U(1)$ gauge theory  sketched  in fig. \ref{tak}, with a bifundamental flavor $A,\At$ and $K$ flavors for each $U(1)$ node, that we denote $P_i, \Pt_i$ and $Q_i, \Qt_i$.

There is also a  cubic superpotential with $2K$ terms:
\be \CW_{\CT_{A,K}}= \sum_{i=1}^K ( A P_i Q_i + \At \Pt_i \Qt_i ) \label{WTA}\,. \ee
The global symmetries of the setup are the non Abelian $SU(K)^2$ acting on the flavors, the two topological $U(1)$ symmetries, and two additional $U(1)$ flavor symmetries. The last $U(1)^2$ flavor symmetry acts exactly as $U(1)_t$ and $U(1)_\phi$ in $\CT_A$.

The exact superconformal R-symmetry mixes with  $U(1)_t$  so it has to be obtained from Z-minimization. We parameterize the R-symmetry in terms of the R-charge of the $4K$ flavor fields, that we denote $r$ and leave $r$ as an unknown parameter. All the flavor fields have the same R-charge as a consequence of the $SU(K)^2$ global symmetry and of an additional global $\IZ_2 \times \IZ_2$ symmetry.

From the exact marginality of the superpotential (\ref{WTA}), it follows that the $2$ bifundamental chirals $A,\At$ have R-charge $2-2r$.

The global charges of the monopole operators can be obtained by means of (\ref{MONDelta}). Denoting by  $\M_{\CT_{A,K}}^{a,b}$ the monopole with topological charges $(a,b)$, we find
 \be R[\M_{\CT_{A,K}}^{a,b}] = K(|a|+|b|)(1-r)+|a-b|(1-(2-2r)) \ee
 
 Notice that for $K>1$ $R[\M_{\CT_{A,K}}^{1,1}] \neq R[\M_{\CT_{A,K}}^{1,0}]$. As usual, a generic monopole operator in the chiral ring is equivalent to a product of the basic monopole operators $\M_{\CT_{A,K}}^{\pm1,0}, \M_{\CT_{A,K}}^{0,\pm1},\M_{\CT_{A,K}}^{\pm(1,1)}$, for instance if $a>b>0$ we have:
 \be R[\M_{\CT_{A,K}}^{a,b}] = (a-b) R[\M_{\CT_{A,K}}^{1,0}] + b R[\M_{\CT_{A,K}}^{1,1}] \,. \ee

Concluding, the chiral ring of $\CT_{A, K}$ is generated by 
\begin{itemize}
\item[-] $K^2+K^2$ mesons $P_i\Pt_j$ and $Q_i\Qt_j$, with R charge $2r$
\item[-] $1$ meson $A\At$, with R charge $4-4r$
\item[-] $4$ monopole operators $\M_{\CT_{A,K}}^{\pm1,0}, \M_{\CT_A}^{0,\pm1}$, with R charge $K(1-r)+2r-1$
\item[-] $2$ monopole operators $\M_{\CT_{A,K}}^{\pm(1,1)}$, with R charge $2K(1-r)$\,.
\end{itemize}

Notice that in the $K=2$ case there is a degeneracy: the meson $A\At$ and the $2$ monopole operators $\M_{\CT_{A,2}}^{\pm(1,1)}$ have the same R charge. This suggests an enhanced symmetry. Indeed the $5d$ CFT associated to the pq-web $(2_{D5'} ,  1_{NS5})$ is $U(1)$ with $2$ flavors, which is actually a free theory of $4$ hypers, $[2]-[2]$, and supports the tri-fundamental symmetry $SU(2)^3$. 

\subsection*{Dualization to $\CT_{B,K}$}
\begin{figure}[h]
\centering
\includegraphics[width=3.3in]{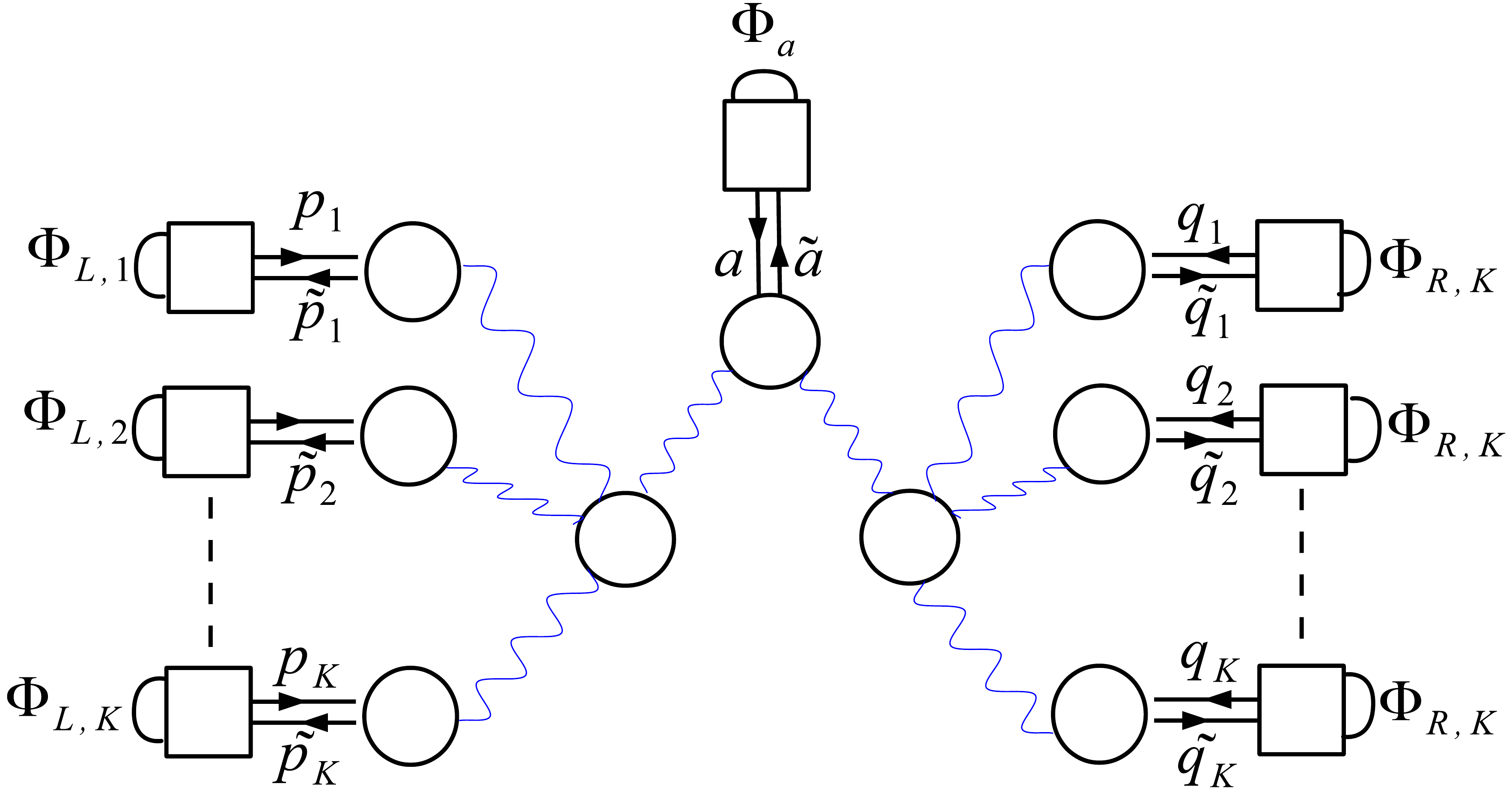}
\caption{The theory $\widehat{\CT}_{A,K}$. Wavy lines  indicate BF couplings.}
\label{bfpic}
\end{figure}

We now group the chiral fields of theory $\CT_{A,K}$ in $2K+1$ pairs of chirals with opposite gauge charges, and dualize each pair using the standard abelian duality
\be \nn \{ P_i, \Pt_i \}, \CW=0 \rightarrow  U(1)_{L,i},{p_i,\pt_i}, \CW=p_i\pt_i\Phi_{L,i}   \ee
\be \label{MS4'} \{ Q_i, \Qt_i \}, \CW=0 \rightarrow  U(1)_{R,i}, {q_i,\qt_i}, \CW=q_i\qt_i\Phi_{R,i}   \ee
\be \nn \{ A, \At \}, \CW=0 \rightarrow  U(1)_a, {a,\at}, \CW=a\at\Phi_{a} \,,  \ee
where the new fields $q_i,\qt_i,p_i,\pt_i$ and $a,\at$ have R charge $1-r$ and $1-(2-2r)$ respectively. In this new dual frame we find 
$\widehat{\CT}_{A,K}$ a  $U(1)^{2K+3}$ theory with $4K+2$ charged chiral fields, $2K+1$ singlet chiral fields and BF couplings depicted in fig. \ref{bfpic}.
 The superpotential contains the $2K+1$ flipping  terms, plus the original $2K$ terms of (\ref{WTA}), which are now written in terms of the monopole operators $\M^\pm_a$ for the   $U(1)_a, {a,\at}$ node,
 $\M^\pm_{i, L}$ for the  $U(1)_{L,i},{p_i,\pt_i}$ nodes and  $\M^\pm_{i, R}$ for the  $U(1)_{R,i}, {q_i,\qt_i}$ nodes: 
 \be \CW_{\widehat{\CT}_{A,K}} = \sum_i (\M^+_a \M^+_{i, L} \M^+_{i, R} + \M^-_a \M^-_{i, L} \M^-_{i, R} + q_{i}\qt_i \Phi_{i, L} + p_{i}\pt_i \Phi_{i, R}) + a\at \Phi_a\,.
\label{bfstep}
\ee

\begin{figure}[h]
\centering
\includegraphics[width=5in]{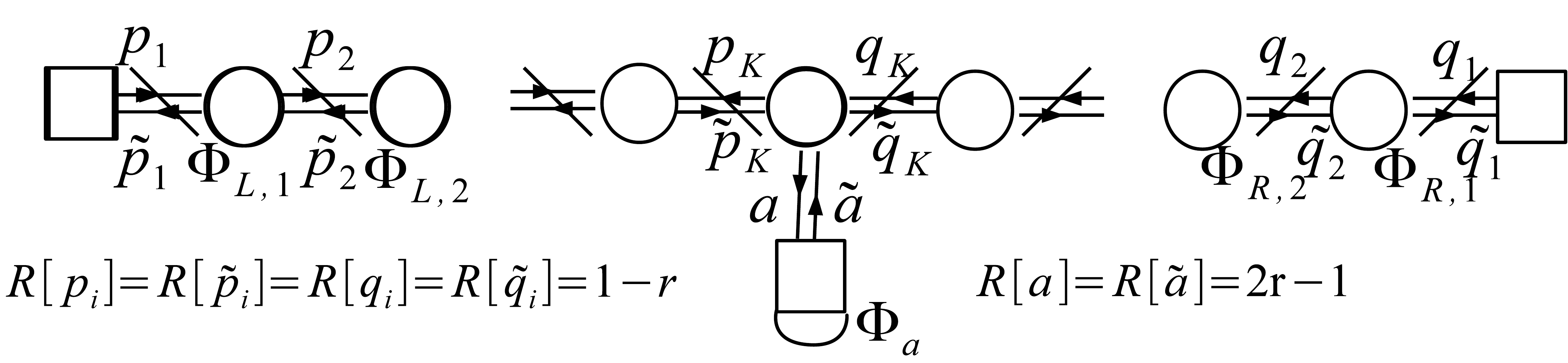}
\caption{The  theory $\CT_{B, K}$.}
\label{tbk}
\end{figure}

At this point we can integrate over the two original gauge $U(1)$'s, that have zero flavors, and produce a functional Dirac delta. We can implement the Dirac delta to present our gauge theory as the linear quiver $\CT_{B, K}$ depicted in  fig. \ref{tbk}, with $U(1)^{2K-1}$ gauge symmetry.  This theory provides a low energy description of the brane set up (\ref{sbsettak}). 

Let us compute the R charge of some basic monopole operators, recalling that the $4K$ `horizontal' bifundamentals have R charges $1-r$ and the $2$ `vertical' bifundamentals $a,\at$ have R charges $2r-1$.

We find that there are $2K$ monopole operators with topological charge given by a string of $\pm 1$ extending symmetrically from the central node,
 with R charge $2$:
\be \label{monopinW} R[\M^{\pm(0,\ldots,0,1,\ldots,1|1|1,\ldots,1,0,\ldots,0)}] = 1-(1-r) + 1-(1-r) +1-(2r-1) = 2 \ee
The part of the superpotential involving the monopoles, which is the sum of $2K$ terms, is precisely the sum of the $2K$ operator in eq. (\ref{monopinW}). These operators are not in the chiral ring.

There also $4\binom{K}{2}=2K(K-1)$ monopole operators with topological charge given by a string of $\pm1$ extending
to the right or to the left of the  central node,
 with R charge $2r$:
\be R[\M^{\pm(0,\ldots,0,1,\ldots,1,0,\ldots,0|0|0,\ldots,0)}] = R[\M^{\pm(0,\ldots,0|0|0,\ldots,0,1,\ldots,1,0,\ldots,0)}] =1-(1-r) + 1-(1-r)  = 2r \ee
These operators combine with the $2K$ singlets $\Phi_{i, L}$ and $\Phi_{i, R}$ to form $2$ $SU(K)\times SU(K)$ bifundamental representation of R charge $2r$. These are dual to the $2K^2$ mesons $P_i\Pt_j$ and $Q_i\Qt_j$ of theory $\M_{\CT_A}^{a,b}$.

The chiral ring of $\CT_{B, K}$ is generated by 
\begin{itemize}
\item[-] $K^2+K^2$ monopole and singlets $\Phi_{i, L/R}$ operators, with R charge $2r$.
\item[-] $1$ singlet $\Phi_a$, with R charge $4-4r$
\item[-] $4$ mesons $\prod_{i=1}^K p_i a$, $\prod_{i=1}^K \pt_i \at$, $\prod_{i=1}^K q_i \at$, $\prod_{i=1}^K \qt_i a$, with R charge $K(1-r)+2r-1$
\item[-] $2$ mesons $\prod_{i=1}^K p_i \prod_{i=1}^K \qt_i$, $\prod_{i=1}^K \pt_i \prod_{i=1}^K q_i$, with R charge $2K(1-r)$
\end{itemize}
and precisely maps to the generators of the chiral ring of $\CT_{A,K}$.

\subsection*{Dualization to $\CT_{C,K}$}
\begin{figure}[h]
\centering
\includegraphics[width=5in]{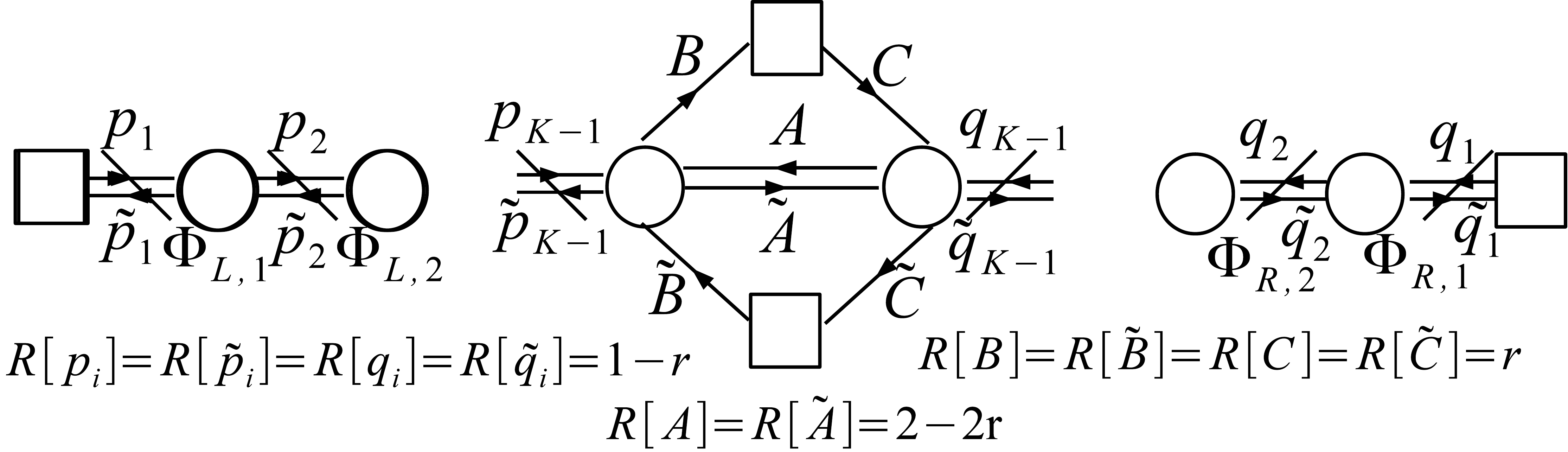}
\caption{ The theory $\CT_{C, K}$}
\label{tck}
\end{figure}

We can also use the triality discussed in detail in Section \ref{triality} to give a different presentation of the theory associated to the brane setup  (\ref{sbsettak}). 
Inside $\CT_{B,K}$ we can see a subquiver isomorphic to $\CT_B$: $U(1)$ with $3$ flavors ($a,\at$ with $R=2r-1$ and $p_K,\pt_K; q_K,\qt_K$ with $R=1-r$) plus $3$ flipping fields ($\Phi_a, \Phi_{K,L}, \Phi_{K,R}$) and the monopole superpotential $ \M^{(0,\ldots,0|+1|0,\ldots,0)}+ \M^{(0,\ldots,0|-1|0,\ldots,0)}$). Using the duality between $\CT_B$  and $\CT_C$ we can locally modify the quiver of fig. \ref{tbk} to the quiver of fig. \ref{tck}, replacing $\CT_B$ with a copy of $\CT_C$, i.e. a pair of XYZ models with fields that we label $A,\At$ with $R=2-2r$ and $B,\Bt; C\Ct$ with $R=r$. Theory $\CT_{C,K}$ has $2K-2$ nodes and the superpotential now reads
\be \CW_{T_{C,K}} = \CW_{flip} + \CW_{monopole} + A B C + \At \Bt \Ct\,, \ee
where both
\be \CW_{flip} = \sum_{i=1}^{K-1} ( q_{i}\qt_i \Phi_{i, L} + p_{i}\pt_i \Phi_{i, R}) \ee
and
\ben  \CW_{monopole} &=& \M^{+(0,0,\ldots,0,1|1,0,\ldots,0,0)} + \nn \\
   && \M^{+(0,\ldots,0,1,1|1,1,0,\ldots,0)} +  \\
    && \;\;\; \ldots\ldots\ldots \;\;\; \nn  + \\
   && \M^{+(1,1,\ldots,1,1|1,1,\ldots,1,1)} \nn + \\
   && \M^{-(0,0,\ldots,0,1|1,0,\ldots,0,0)} + \nn \\
   && \M^{-(0,\ldots,0,1,1|1,1,0,\ldots,0)} + \nn  \\
    && \;\;\; \ldots\ldots\ldots \;\;\; \nn  + \\
   && \M^{-(1,1,\ldots,1,1|1,1,\ldots,1,1)} \nn \een
have $2K-2$ terms.

It is easy to check that the chiral ring generators are precisely the same as in $\CT_{B,K}$. One non-trivial aspect of this theory is that the two central $U(1)$, are `balanced' even if they support $3$ flavors instead of $2$. Let us check this statement computing the scaling dimension of the simple monopole associated to those nodes: $\M^{(0,\ldots,0,1|0,\ldots,0)}$.
\be R[\M^{(0,\ldots,0,1|0,\ldots,0)}] =  1-r_{q_{K-1}} + 1 - r_A + 1-r_B = (r) + (-1+2r) + (1 -r) = 2r\ee
which is equal to the scaling dimension of the monopoles associated to the other nodes of the quiver with $2$ flavors. 

\subsection*{S-duality and gauge theory tetrality}
\begin{figure}[h]
\centering
\includegraphics[width=5in]{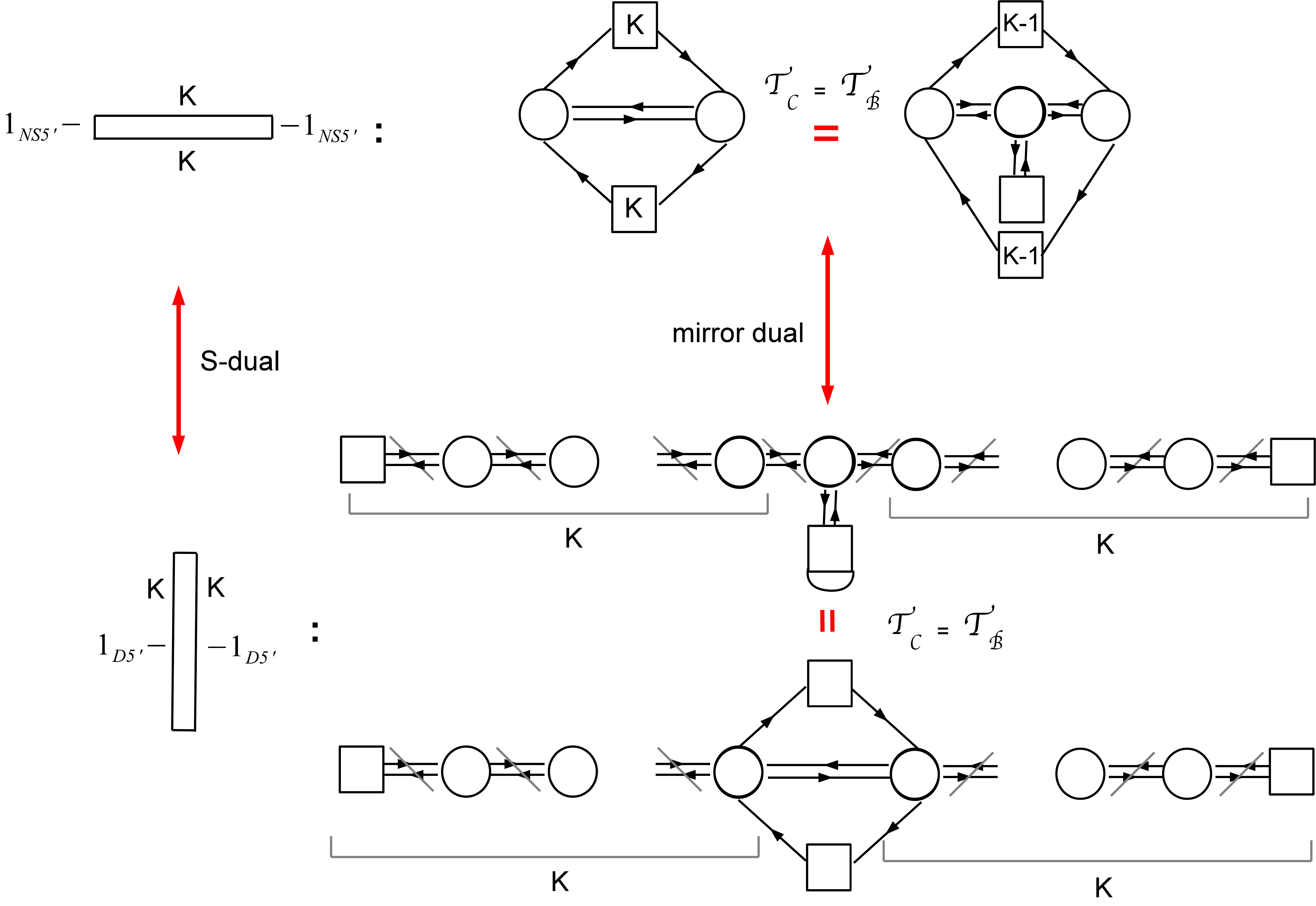}
\caption{A gauge theory tetrality for two S-dual brane set ups. The quiver theories in the first line are related by the application
of the  $\CT_C \rightarrow \CT_B$ duality and describe the $(1_{NS5}, K_{D5'})$-web.  
The quiver theories in the third and forth lines  describe  the S-dual  $(K_{NS5}, 1_{D5'})$-web.  
 They are  related to each other by the application
of the  $\CT_C \rightarrow \CT_B$ duality and are obtained from those on the first line by piece-wise mirror-dualisation. }
\label{tetrak}
\end{figure}

We can also apply the $\CT_C \rightarrow \CT_B$ replacement  inside theory $\CT_{A,K}$. We get a new theory
$\CT_{D,K}$,  a $U(1)^3$ quiver gauge theory with $K-1$ flavors for the external nodes and $1$ flavor for the central node,
depicted in the top right corner of fig. \ref{tetrak}.
Here the obvious global symmetry is only $SU(K-1)^2 \times U(1)^6$, but there must be an enhancement to $SU(K)^2 \times U(1)^4$.

Naming the bifundamental fields $a,\at$, $b,\bt$ and the central flavor $c,\ct$, the superpotential now reads
\be \CW_{\CT_{D,K}}= \sum_{i=1}^{K-1} ( a b P_i Q_i + \at \bt \Pt_i \Qt_i ) + \M^{(0,1,0)} + \M^{(0,-1,0)} + \Phi_a a\at + \Phi_b b \bt + \Phi_c c\ct \label{WTA'} \,.\ee

Again, it's easy to check that the chiral ring generators are the same. For instance the left $K^2$ mesons $P_i \Pt_j \, i,j=1 \ldots K$ are now given by 
\begin{itemize}
\item[-] the $(K-1)^2$ quadratic operators $P_i \Pt_j. \, i,j=1, \ldots,  K-1$
\item[-] the $2(K-1)$ cubic operators $P_i \at c$ and $\Pt_i a \ct$, $i=1, \ldots,  K-1$
\item[-] the quartic operator $a\at c\ct$
\end{itemize}

Notice that  in this case the global symmetry enhancement is not of type $(U(1)_{top} \times U(1)_{axial})^{K-1} \rightarrow SU(K)^2$. Here we have two standard mesonic symmetries that combine to give $SU(K-1)\times U(1) \rightarrow SU(K)$. Again, we have no rigorous proof of the enhancement without invoking dualities. It would be nice to compute the Hilbert Series of the chiral rings using the monopole formulae of \cite{Cremonesi:2013lqa}\cite{Cremonesi:2014kwa}\cite{Cremonesi:2014vla}\cite{Hanany:2015via}\cite{Cremonesi:2015dja}.

In conclusion we have shown that there are four dual gauge theories  descriptions for the two S-dual brane set-up as summarized  in fig.  \ref{tetrak}.

\subsection{$S^3$ partition function derivation}

Finally we check the duality web at the level of partition functions.
We start from $\CT_{A,K}$ with real masses $m_a$,  $\sum_{a}^K  m_a=0$, $\phi_a$, $\sum_{a}^K \phi_a=\Phi$ 
in the Cartan of $SU(K)_{m}\times SU(K)_{\phi} \times U(1)_\Phi
 \times U(1)_{t}\times U(1)_{\xi_1}\times U(1)_{\xi_2}$:
 \ben
\label{TAK}
\!\!\!\!\!\!\!\! Z_{\CT_{A,K}}\!\!=\!\! \int dx_1 dx_2 e^{2\pi i (\xi_1 x_1+\xi_2 x_2)}  F_{2t}(x_2-x_1)\prod_{i=1}^K  F_{2v-\phi_a}(x_1+m_a) F_{2v+\phi_a}(x_2+m_a)\,.
\een

To obtain the mirror dual we first use eq. (\ref{fund}) to dualize the  bifundamentals, we then collect 
all the fields connected to the $x_1$ node and  use eq. (\ref{linmir}) to dualize  into the  mirror linear quiver:
\ben
\int dx_1 e^{2\pi i x_1(\xi_1+s)}\prod_{a=1}^K F_{2v-\phi_a}(x_1+m_a)
 =\prod_{a=1}^K s_b(2v-\phi_a) \int  \prod_{a=1}^{K-1}  dw_a  ~T_{K}(\vec w,\vec m, 2v-\vec \phi, \xi_1+s)\,.\nn\\
 \een
We do the same for the $x_2$ integration 
\be
\int dx_2 e^{2\pi i x_2(\xi_2-s)}\prod_{a=1}^K F_{2v+\phi_a}(x_2+m_a)=
\prod_{a=1}^K s_b(2v+\phi_a) \int  \prod_{a=1}^{K-1}  du_a  ~T_{K}(\vec u,\vec m, -2v-\vec \phi, \xi_2-s)\,,
\ee
and we find
\ben
&(\ref{TAK})=&s_b(2t) \prod_{a=1}^K s_b(2v\pm\phi_a) \int ds F_{-2t}(s)~\nn
\\
&&\times \int  \prod_{a=1}^{K-1}  dw_a~ T_{K}(\vec w,\vec m, -2v+\vec \phi, \xi_1+s)
\nn
\\
&&\times
\int  \prod_{a=1}^{K-1}  du_a  ~T_{K}(\vec u,\vec m, -2v-\vec \phi,\xi_2- s)\,.
\een
We can rewrite this  by shifting $w_a\to w_a+\xi_1$ and $u_a\to u_a+\xi_2$  and sending  $u_a\to -u_a$ 
as:
\ben
&&\!\!\!\!\!\!\!\! \!\!\!=e^{-2\pi i m_1(\xi_1+\xi_2)}  s_b(2t) \prod_{a=1}^K s_b(2v\pm\phi_a) \int ds 
 \int  \prod_{a=1}^{K-1}  dw_a\int  \prod_{a=1}^{K-1}  du_a~
\nn\\
&&\!\!\!\!\!\!\!\!\times
e^{2\pi i  (m_2-m_1) (w_1-u_1)}e^{2\pi i   (m_3-m_2) (w_2-u_2)} \cdots e^{2\pi i (m_K-m_{K-1}) (w_{K-1} - u_{K-1} )}
\nn\\
&&\!\!\!\!\!\!\!\!\times
 F_{-2t}(s)
F_{-2v+\phi_K}(w_{K-1}-s)F_{2-v-\phi_K}(u_{K-1}-s)\nn\\
&&\!\!\!\!\!\!\!\!\times
F_{-2v+\phi_1} (w_1+\xi_1)
F_{-2v+\phi_2} (w_2-w_1) \cdots F_{-2v+\phi_{K-1}} (w_{K-1}-w_{K-2}) 
\nn\\
&&\!\!\!\!\!\!\!\!\times
F_{-2v-\phi_1} (u_1-\xi_2)
F_{-2v-\phi_2} (u_2-u_1) \cdots F_{-2v-\phi_{K-1}} (u_{K-1}-u_{K-2})=Z_{\CT_{B,K}}\,.
\label{TBK}
\een
We observe that  the real masses $\phi_a$ on the l.h.s. and on the r.h.s.  are identified and that there is  no real mass turned on for the central $U(1)_{top,s}$. This is  consistent with  $\CW_{\CT_{B,K}}$ breaking the global symmetries on the two tails to the diagonal and the central node topological symmetry. The $2K+1$ flipping terms are also visible.  
This allows us to identify (\ref{TBK})  with  the partition function of  $\CT_{B,K}$.

We can integrate over the central node by using the new pentagon identity (\ref{newpent}):
%
\ben
&&\!\!\!\!\!\!\!\!\!\!\!=e^{-2\pi i m_1(\xi_1+\xi_2)}   \prod_{a=1}^{K-1}s_b(2v\pm\phi_a) \int  \prod_{a=1}^{K-1}  dw_a\int  \prod_{a=1}^{K-1}  du_a~
\nn\\
&&\!\!\!\!\!\!\! \times
e^{2\pi i  (m_2-m_1) (w_1-u_1)}e^{2\pi i   (m_3-m_2) (w_2-u_2)} \cdots e^{2\pi i (m_K-m_{K-1}) (w_{K-1} - u_{K-1} )}
\nn\\
&&\!\!\!\!\!\!\!\times
F_{2t}(w_{K-1}-u_{K-1})
F_{2v+\phi_K}(w_{K-1})F_{2v-\phi_K}(u_{K-1})\nn\\
&&\!\!\!\!\!\!\!\times
F_{-2v+\phi_1} (w_1+\xi_1)
F_{-2v+\phi_2} (w_2-w_1) \cdots F_{-2v+\phi_{K-1}} (w_{K-1}-w_{K-2}) 
\nn\\
&&\!\!\!\!\!\!\!\times
F_{-2v-\phi_1} (u_1-\xi_2)
F_{-2v-\phi_2} (u_2-u_1) \cdots F_{-2v-\phi_{K-1}} (u_{K-1}-u_{K-2})=Z_{\CT_{C,K}} \,.
\label{TCK}
\een
Again we observe that the  real masses on the left and right  tails are identified which  is consistent with 
$\CW_{\CT_{C,K}}$. There are also the  $2K-2$ flipping terms hence we can identify (\ref{TCK}) as the  partition function of  $\CT_{C,K}$.

Finally we find the last  dual presentation to complete the diagram in fig.  \ref{tetrak}. 
In (\ref{TAK}) we  implement the $\CT_C=\CT_B$ duality
to replace:
\ben
\nn
&&F_{2t}(x_2-x_1)  F_{2v-\phi_K}(x_1+m_K) F_{2v+\phi_K}(x_2+m_K)=\\
&&
s_b(2t)     s_b(2v\pm \phi_K) \int ds~  F_{-2t}(s)  F_{-2v-\phi_K}(s+x_1+m_K) F_{-2v+\phi_K}(s+x_2+m_K)\,.
\een
If substitute this into eq.  (\ref{TAK}) we  find
\ben\nn
&&= s_b(2t)     s_b(2v\pm \phi_K) 
\int dx_1 dx_2 e^{2\pi i (\xi_1 x_1+\xi_2 x_2)}  \prod_{a=1}^{K-1}  F_{2v-\phi_a}(x_1+m_a) F_{2v+\phi_a}(x_2+m_a)\\
&&\times
 \int ds~  F_{-2t}(s-m_K)  F_{-2v-\phi_K}(s+x_1) F_{-2v+\phi_K}(s+x_2)=Z_{\CT_{D,K}}\,,
\een
which is the partition function of $\CT_{D,K}$. From this expression we see  how the real masses are mapped and conclude that indeed as anticipated the symmetry enhancement involves only mesonic symmetries.

\section{Massive deformations}\label{deform}

\subsection*{Complex mass deformations}
\begin{figure}[h]
\centering
\includegraphics[width=6in]{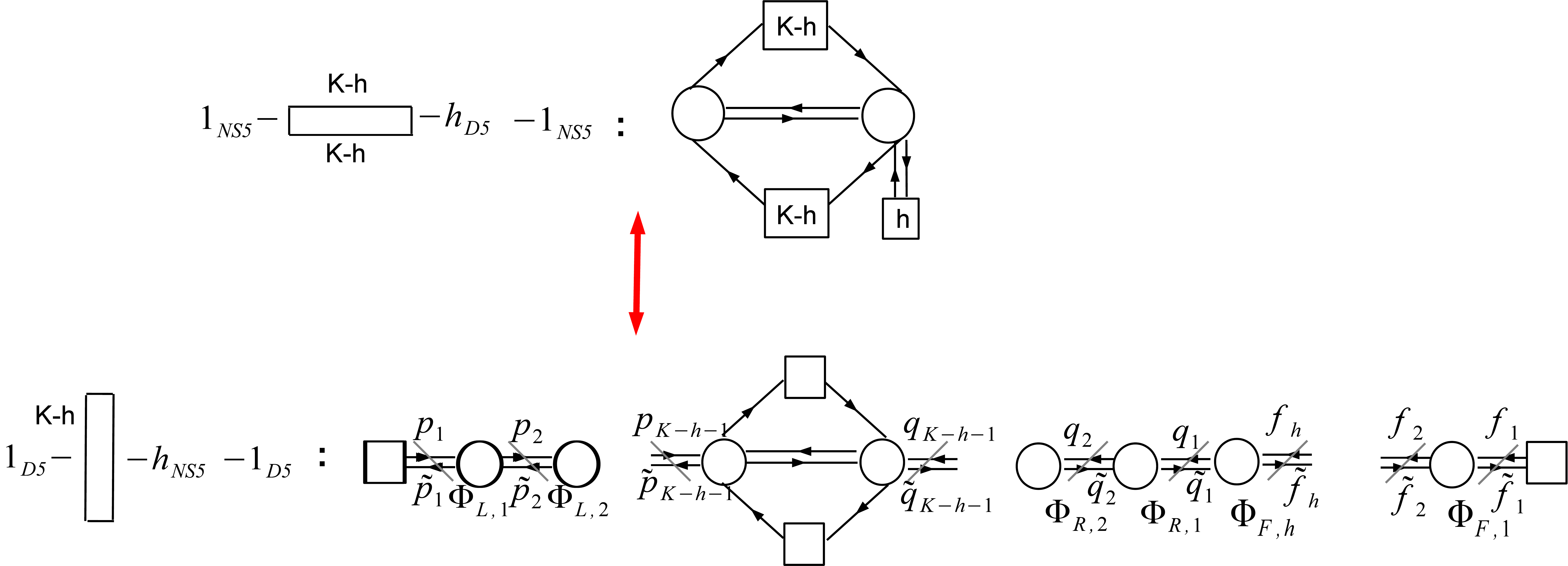}
\caption{The effect of a complex mass deformation.}
\label{comass}
\end{figure}

Consider deforming  the $1_{NS5} - (K_{D5} ,  1_{NS5'})  - 1_{NS5}$ set-up to  $1_{NS5}- (K-h_{D5} ,  1_{NS5'})  - h_{D5}  - 1_{NS5}$. We are extracting $h$ $D5$  from the pq-web and pushing them to the right. This deformation corresponds in the 3d  low energy theory  to turning on  a complex  mass for $h$  mesons in $\CT_{A,K}$:
\be\CW_{\CT_{A,K}} \rightarrow \CW' = m \sum_{i=1}^h P_i\Pt_i + \sum_{i=1}^K ( A P_i Q_i + \At \Pt_i \Qt_i)\,. \ee
Integrating out the $2h$ massive fields $P_i, \Pt_i\,,\,i=1,\ldots,h$ we obtain at low energies the theory
$\CT^{def}_{A,K}$  depicted in fig. \ref{comass}, a $U(1)^2$ quiver gauge theory with $4(K-h)$ chirals entering a cubic superpotential and $2h$ chirals entering a quartic superpotential:
\be \CW_{\CT^{def}_{A,K}}= - \frac{1}{m} A \At \sum_{i=1}^h Q_i \Qt_i + \sum_{i=h+1}^K ( A P_i Q_i + \At \Pt_i \Qt_i) \label{cub+quart}\,.\ee
This superpotential supports a global symmetry $SU(K-h)^2 \times U(1) \times U(h) \times U(1)^2_{top} \times U(1)_t$, as expected from the global symmetry of the brane system. Notice that the global symmetry associated to the stack of $h$ $D5$ branes is now $U(h)$ rather than $SU(h)^2 \times U(1)$, since the $k$ $D5$'s are not broken into two parts. 
The complex mass parameter $m$ is related to the distance between the $(K-h_{D5'},1_{NS})$ pq-web and the $h$-D5's stack. When this distance goes to zero, there is a divergence in the Lagrangian that must be cured integrating in the $2h$ chiral fields $P_i, \Pt_i$.\footnote{Notice that in the brane setup $1_{NS5'} - N_{D3} - h_{D5}  - N_{D3} - 1_{NS5'}$ the non Abelian global symmetry is $SU(h)^2$, while in the brane setup $1_{NS5'} - N_{D3} - 1_{NS5'} - M_{D3} - h_{D5}  - M_{D3} - 1_{NS5'}$ the non Abelian global symmetry is $SU(h)$. In the latter case the quartic superpotential term involving the $h$ flavors and the bifundamentals breaks $SU(h)^2\rightarrow SU(h)$. In the former case it's impossible to write such a superpotential term so there is an accidental chiral symmetry in the gauge theory. We are grateful to Amihay Hanany for discussions about this point.}

As an aside, if we push $h_R$ $D5'$s to the right and $h_L$ $D5'$s to the left of the pq-web, while $K-h_L-h_R$ $D5$'s remain on top of the $NS5'$, we obtain a $[K-h_R]-(1)-(1)-[K-h_L]$ quiver with superpotential \be \CW= A\At ( \sum_{i=1}^{h_L} P_i \Pt_i + \sum_{i=1}^{h_R} Q_i \Qt_i  ) +  \sum_{i=h_L+h_R+1}^{K}  ( A P_i Q_i + \At \Pt_i \Qt_i )\,.  \ee

Let us go back to the theory $\CT^{def}_{A,K}$ and try to determine the mirror dual. We can 
start from the  dual $(1_{D5'}, K_{NS})$ web and extract $h$ $NS5$ branes. This leads to the  low energy theory 
$\CT^{def}_{C,K}$ which is  given by $\CT_{C,K-h}$, with the monopole superpotential mirror of the cubic part of (\ref{cub+quart}), joined to an additional  $U(1)^h$ tail as shown in fig. \ref{comass}.
 The Coulomb branch operators (adjoint singlets $\Phi_i$ and monopoles) of these additional $h$ nodes enter the superpotential through the mirror of the quartic part in (\ref{cub+quart}), $A \At (\sum_i Q_i \Qt_i )$. Recalling the map (\ref{basicmap}), we see that whole superpotential is given by
\be \CW_{\CT^{def}_{C,K}}= \CW_{flip} + \CW_{monopole} - \frac{1}{m} a \at ( \sum_{i=1}^h \Phi_i) \,.\ee
The adjoint singlets couple to the central bifundamentals $a,\at$ through the coupling   $a \at ( \sum_{i=1}^h \Phi_i)$ which breaks  all the axial symmetries of the $U(1)^h$ tail, the $U(1)^h$ topological symmetries instead enhance to $U(h)$. Notice also that the coupling  $a \at ( \sum_{i=1}^h \Phi_i)$ is non-local in the quiver: the $h$ singlets $\Phi_i$ enter the superpotential both with a bifundamental in the right $U(1)^h$ tail, and with the bifundamental connecting the two $U(1)^{K-h-1}$ tails.

We can check the duality at the level of partition function.
For $\CT^{def}_{A,K}$ we have:
\ben
&& Z_{\CT^{def}_{A,K}}=\int dx_1 dx_2 e^{2\pi i (\xi_1 x_1+\xi_2 x_2)}  F_{2t}(x_2-x_1)\nn\\
&&\times \prod_{i=1}^{K-h}  F_{2v-\phi_a}(x_1+m_a) F_{2v+\phi_a}(x_2+m_a)
\prod_{j=1}^h F_{2v+\zeta}(x_2+M_j)\,,
\een
where the  quartic superpotential   imposes  $\zeta=2v$.

It is then easy to check that the  piecewise dualisation leads to:
\ben
&&\!\!\!\!\!\!\!\!\!\!\!
Z^{def}_{\CT_{C,K}} 
=e^{-2\pi i (m_1\xi_1+M_1\xi_2)}  
s_b(4v)^h \prod_{a=1}^{K-h-1}s_b(2v\pm\phi_a) \int  \prod_{a=1}^{K-h-1}  dw_a\int  \prod_{a=1}^{K-1}  du_a~
\nn\\
&&\!\!\!\!\!\!\! \times
e^{-2\pi i  u_1(M_2-M_1)}\cdots e^{-2\pi i  u_{h-1}(M_h-M_{h-1})}
\nn\\
&&\!\!\!\!\!\!\! \times
e^{2\pi i  (m_2-m_1) (w_1-u_{h+1})}e^{2\pi i   (m_3-m_2) (w_2-u_{h+2})} \cdots e^{2\pi i (m_{K-h}-m_{K-h-1}) (w_{K-h-1} - u_{K-1} )}
\nn\\
&&\!\!\!\!\!\!\!\times
F_{2t}(w_{K-h-1}-u_{K-1})
F_{2v+\phi_{K-h}}(w_{K-h-1})F_{2v-\phi_{K-h}}(u_{K-1}  )\nn\\
&&\!\!\!\!\!\!\!\times
F_{-2v+\phi_1} (w_1+\xi_1)
F_{-2v+\phi_2} (w_2-w_1) \cdots F_{-2v+\phi_{K-h-1}} (w_{K-h-1}-w_{K-h-2}) 
\nn\\
&&\!\!\!\!\!\!\!\times
F_{-2v-\phi_1} (u_{h+1}-u_h)
\cdots 
F_{-2v-\phi_{K-h-1}}(u_{K-h-1}-u_{K-h-2})
\nn\\ 
&& \!\!\!\!\!\!\! \times
 F_{-4v}(u_1-\xi_2) F_{-4v}(u_2-u_1)\cdots F_{-4v}(u_h-u_{h-1})
 \,.
\een
Notice that the effect of the coupling $a \at ( \sum_{i=1}^h \Phi_i)$  which breaks the axial symmetries of the joint $U(1)^h$ tail is visible in the last line where all the axial masses are set equal to $-2v$.

\subsection*{Real mass deformations}

\begin{figure}[h]
\centering
\includegraphics[width=6in]{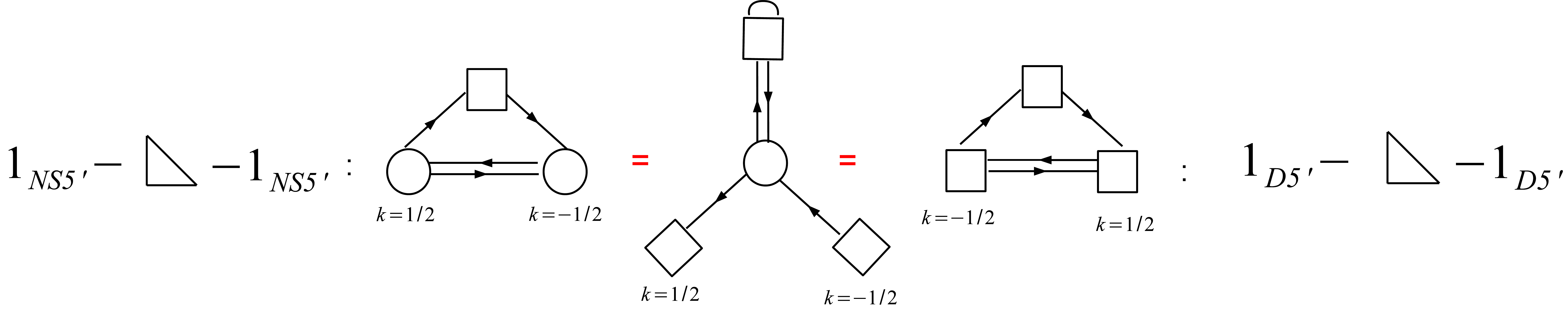}
\caption{The effect of a real mass deformation on the triality. }
\label{fig11realmass}
\end{figure}

A different  type of deformations of our brane set-ups  correspond to turning on real masses in the 3d low energy the theories.
For example, in the brane set-up for $\CT_A$  in fig. \ref{bset}, we consider the deformation where we move toward infinity along $x_7$  one of the two $D5'$ halves. The $(1_{NS5},1_{D5'})$-web is deformed  to a junction of three  5-branes. The toric diagram associated to the junction  is the triangle with unit base and hight shown fig.   \ref{fig11realmass}.
This deformation  corresponds to the following   limit on the real masses:\footnote{With  different choices of signs we  can  integrate out chirals of opposite charge generating  CS terms with opposite levels, see for example \cite{Cremonesi:2010ae}.
Here we are only interested in a qualitative analysis of the  effect of these real mass deformations on the dual theories.}
\be
\label{11real}
\phi=2s \,, \qquad \xi_{1,2}\to \xi_{1,2}\pm s \, \qquad s\to \infty.
\ee
The key property we use is the asymptotic of the double sine function:
\be
\label{asym}
\lim_{x\to \pm  \infty}s_b(x)\sim e^{\pm i\pi \frac{x^2}{2}}\,.
\ee
In $Z_{\CT_A}$ we need to ``follow the vacuum" by shifting $x_{1,2} \to x_{1,2}+s$ to find
\ben
\nn&& \lim_{s\to\infty}Z_{\CT_A} = e^{i\pi s(4v+i Q)}~
\int dx_1 dx_2 ~
e^{\tfrac{i\pi}{2} (\tfrac{iQ}{4} +v+x_2)^2}e^{-\tfrac{i\pi }{2}(\tfrac{iQ}{4} +v-x_1)^2}
\\
&&~~\times e^{2\pi i (x_1\xi_1+x_2\xi_2)}
F_{2t}(x_2-x_1)  s_b(\tfrac{iQ}{4}+v+x_1)
s_b(\tfrac{iQ}{4}+v-x_2)  \,.
\een
The effect of the limit is to remove one chiral on each node and to replace it with half CS units of opposite sign as shown in the first quiver from  the left in  fig. \ref{fig11realmass}. The superpotential  contains only the cubic term $\tilde P\tilde A \tilde Q$.

In $\CT_B$ the limit (\ref{11real}) yields:
\ben
\nn&& \lim_{s\to\infty}Z_{\CT_B} = e^{i\pi s(4v+i Q)}~s_b(2t)
\int dz~
e^{\tfrac{i\pi}{2} (\tfrac{iQ}{4} -v+z+\xi_1)^2}e^{-\tfrac{i\pi }{2}(\tfrac{iQ}{4} -v-z+\xi_2)^2}
\\
&&~~\times 
F_{-2t}(z)  s_b(\tfrac{iQ}{4}-v+z-\xi_2)
s_b(\tfrac{iQ}{4}-v-z-\xi_1)  \,,
\een
in this case the limit removes two chirals and two singlets and creates some background CS units as shown in the centre of fig. \ref{fig11realmass}.
The superpotential consists of a cubic term  flipping the hypermultiplet and  the linear  monopole  term $\M^+$ to which 
$\tilde P\tilde A \tilde Q$ in $\CT_A$ is mapped.

Finally on in  $\CT_C$ the limit (\ref{11real}) gives:
\ben
&& \lim_{s\to\infty}Z_{\CT_C} = e^{i\pi s(4v+i Q)}
e^{\tfrac{i\pi}{2} (\tfrac{iQ}{4} +v+\xi_1)^2}e^{-\tfrac{i\pi }{2}(\tfrac{iQ}{4} +v+\xi_2)^2}
F_{2t}(\xi_1+\xi_2)  s_b(\tfrac{iQ}{4}+v-\xi_2)
s_b(\tfrac{iQ}{4}+v-\xi_1)  \,.\nn\\
\een
Again two chiral multiplets are removed and replaced by  two background half CS units of opposite sign as shown on last quiver on the right in fig. \ref{fig11realmass}.

\begin{figure}[h]
\centering
\includegraphics[width=6in]{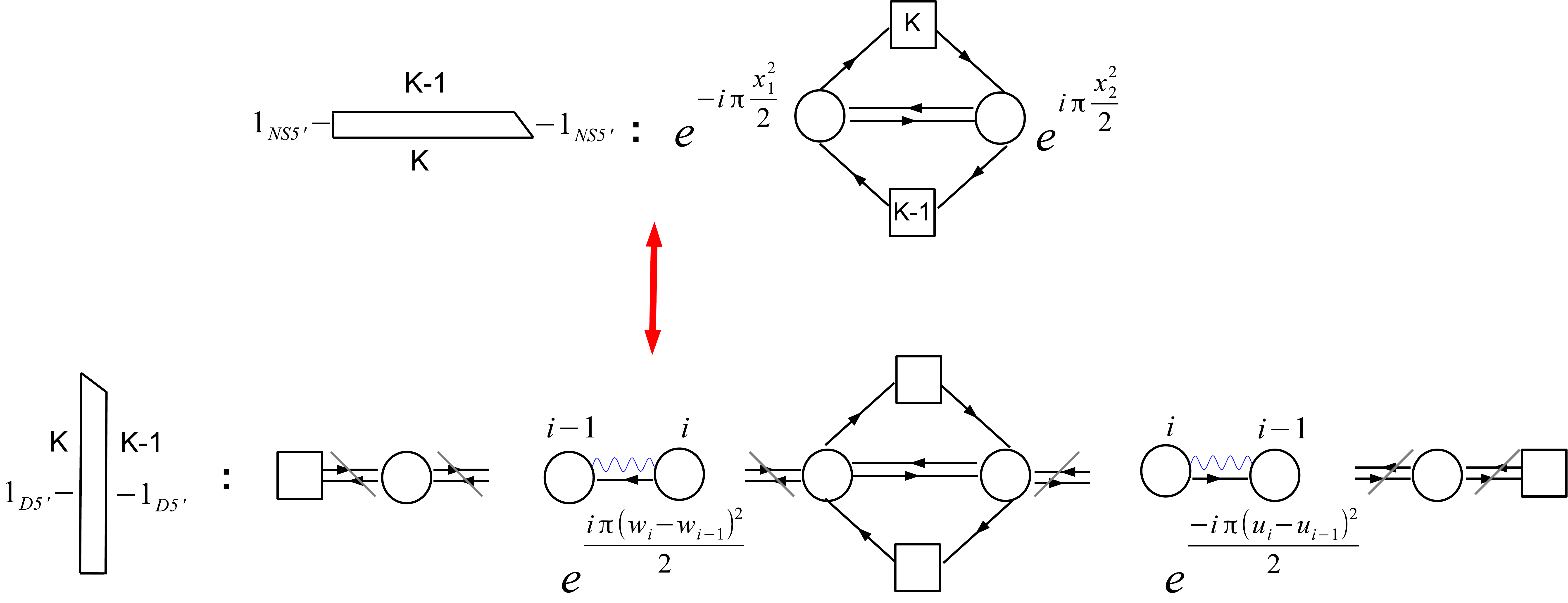}
\caption{ The effect of a chiral mass deformation on $T_{A,K}$ and on its mirror $T_{C,K}$. }
\label{realdef}
\end{figure}

It is easy to  generalize this mass deformation to the case $(1_{NS5}, K_{D5'})$ case. When we send to infinity a $D5'$ half the $(1_{NS5}, K_{D5'})$-web is deformed into the configuration corresponding to a toric diagram given by a right trapezoid with bases of length $K$ and $K-1$ and unit height  shown in fig. \ref{realdef}.
To implement this limit we define a new set of vector masses and axial masses
\be
\phi_a\to\phi_a\,, \quad m_a\to m_a- s\,, \quad a \neq i\,,  \qquad m_i\to m_i+(K-1) s\,, \qquad \phi_i\to2Ks\,, \quad \xi_{1,2}\to \xi_{1,2}\pm s\,.
\ee
In $\CT_{A,K}$ we  shift the integration variables $x_{1,2}\to x_{1,2}+s$ and use the  asymptotics (\ref{asym})
to find (we also absorb some linear terms in a  renormalisation of the FI parameters):
\ben
\label{mta}
\lim_{s\to \infty}Z_{\CT_{A,K}}&\sim&
\int dx_1 dx_2~ e^{2\pi i (\xi_1 x_1+ \xi_2 x_2)} ~e^{\tfrac{i\pi(x_2^2-x_1^2)}{2}}
\nn\\ &&\times
F_{2t}(x_2-x_1)s_b(\tfrac{iQ}{4}+v+x_1+ m_i)
s_b(\tfrac{iQ}{4}+v-x_2- m_i)\nn\\
&&\times \prod_{a\neq i}^{K} F_{2v- \phi_a}(x_1+m_a)  F_{2v+\phi_a}(x_2+ m_a)\,.
\een
The effect of this limit is to remove one chiral on the left and one on the right and to  introduce two opposite half CS units
as shown  on the top of fig. \ref{realdef}. The superpotential coupling $P_i A Q_i$ is removed.

The mirror dual can be easily obtained via piece-wise dualisation.  We apply  the  pentagon identity (\ref{fund}) to dualize  the $2(K-1)$ hypers connected to the left and right nodes and the bifundamental hyper.
The chirals with half CS unit are instead dualized by means of eq. (\ref{csh}). After implementing the delta functions and applying the  new pentagon identity  (\ref{newpent}) we can present the result as the quiver on the bottom of  fig. \ref{realdef}. On each tail the $i$-th bifundamental  chiral is removed and a BF coupling with  half CS unit is created. 
Finally in the superpotential we need to remove the term  $ \M^{-(0,\ldots,0,1,\ldots1,1|1,1,\ldots,1,0\ldots,0)}$ dual to the cubic coupling   $P_i A Q_i$  containing the chirals that we have integrated out.\\

One can iterate this deformation and move to infinity all the $K$  $D5'$ halves on the same side of the $NS5$ and  reach a configuration described by a toric diagram given by a right triangle of base $K$ and unit hight. It should be clear that on the gauge theory side this deformation  corresponds to integrate out  all the fundamental  chirals  in  the lower triangle in $\CT_{A,K}$. On the dual  side instead we are left with a deformed version of  $\CT_{C,K}$ where the quiver tails contain chiral bifundamentals going only in one direction plus BF couplings.

\subsection*{Vector mass deformation}

Another interesting  non-minimal  deformation corresponds to moving towards  infinity two  $D5'$ halves on opposite side  of the $NS5$ along the  $x_7$ direction with the same sign. This has the simple effect of removing a $D5'$ from the pq-web: $(1_{NS5},K_{D5'})\to (1_{NS5},K-1_{D5'})$.

To implement this deformation in  $Z_{\CT_{A,K}}$ we define a new set of vector masses
$$
m_a=m_a- s\,, \quad a \neq i\,,  \qquad m_i=m_i+(K-1) s\,,
$$
we shift the integration variables $x_{1,2}\to x_{1,2}+s$ and 
take the  $s \to \infty$ limit.  Only the $i^{th}$ left and right  hypers are affected by this limit which we evaluate by using the asymptotics (\ref{asym}):
\be
\lim_{s\to \infty} F_{2v-\phi_i}(x_1+m_i +K s) F_{2v+\phi_i}(x_2+m_i +K s)\sim
 e^{i\pi (x_1+ m_i+Ks) (-\phi_i+2v+\tfrac{iQ}{2})}
 e^{i\pi (x_2+ m_i+Ks) (\phi_i+2v+\tfrac{iQ}{2})}\,.
\ee
We see that this limit has the trivial effect of removing two hypers and contributing a finite shift to the FI parameters. The remaining part of the partition function will have no  $m_i$ and $\phi_i$ dependence hence it will be a function of $2K$ real masses only.
We conclude that as expected  this vector mass deformation reduces $\CT_{A,K}\to \CT_{A,K-1}$. The mirror duals are obviously  $\CT_{B,K-1}$ and $\CT_{C,K-1}$.

\subsection*{Axial mass deformation}

\begin{figure}[h]
\centering
\includegraphics[width=6in]{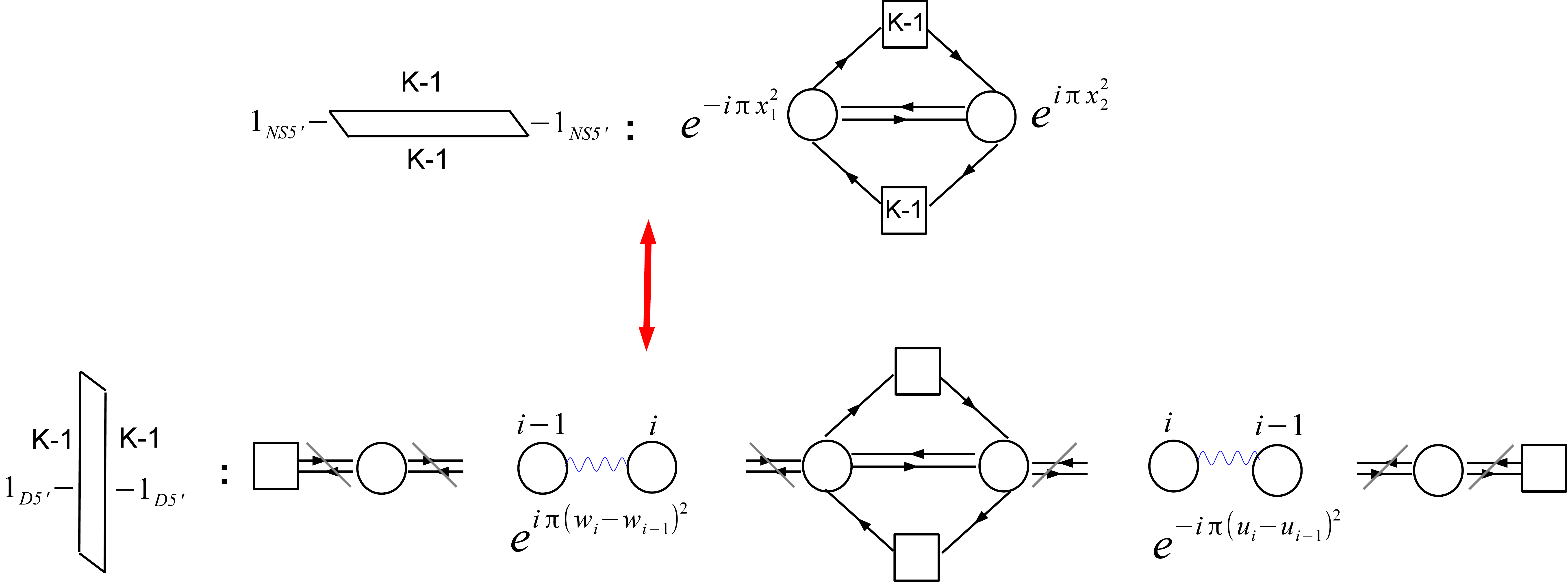}
\caption{ The effect of an axial  mass deformation on $T_{A,K}$ and on its mirror $T_{C,K}$. }
\label{tckaxdef}
\end{figure}

We can also send  two $D5'$-halves, on opposite side of the $NS5$, to infinity along $x_7$ with opposite signs.
In this case the $(1_{NS5'},K_{D5'})$-web  is deformed into the configuration corresponding to a toric diagram given by a parallelogram  with bases of length $K-1$ and unit height  shown in fig. \ref{tckaxdef}.

To implement this deformation we take $\phi_i =s\to \infty$. 
In  $Z_{\CT_{A,K}}$ this limit yields:
\ben
\label{axta}
\lim_{s \to \infty}Z_{\CT_{A,K}}&=&
e^{i \pi s(2v+\tfrac{i Q}{2} )}  
\int dx_1 dx_2 e^{2\pi i (\xi_1 x_1+\xi_2 x_2)} e^{2i\pi m_i(x_2-x_1)  }~F_{2t}(x_2-x_1)
 e^{i\pi (x_2^2- x_1^2)}  
\nn\\&&
\times \prod_{a\neq i}^{K} F_{2v-\phi_a}(x_1+m_a)  F_{2v+\phi_a}(x_2+m_a)\,.
\een
The $i^{th}$ hypers are removed and replaced by  two   Chern-Simons terms with $k=\pm 1$  as  shown in fig. \ref{tckaxdef}.
Notice that actually there is no-dependence on the mass $m_i$ as it can be absorbed in  a shift in the FI parameters.
Also in this case the deformed theory depends on $2K$ real masses only.

In this case the cubic superpotential couplings $P_i A Q_i+\tilde P_i \tilde A \tilde Q_i$ are removed.

We can now find the mirror of this configuration by piecewise dualisation.
We first dualize  all the  hypers by means of the pentagon identity (\ref{fund}).
We then dualize the  Gaussian CS couplings by writing them as inverse  Fourier transforms.
At this point, after few manipulation we can apply the new pentagon identity  (\ref{newpent}) to  present the dual theory as the quiver on the bottom of fig. \ref{tckaxdef} with partition function:
\ben
\label{axtc}
&&
\lim_{s\to \infty}
Z_{\CT_{C,K}} =
e^{i \pi s (2v+\tfrac{i Q}{2} )}
e^{-2\pi i m_1(\xi_1+\xi_2)} 
  \prod_{a\neq i}^{K-1}s_b(2v\pm\phi_a) \int  \prod_{a=1}^{K-1}  dw_a\int  \prod_{a=1}^{K-1}  du_a~
\nn\\
&&\!\!\!\!\!\!\! \times
e^{2\pi i  (m_2-m_1) (w_1-u_1)}
\cdots e^{2\pi i ( w_{i-1}-u_{i-1})m_{i-1}} e^{2\pi i  (u_{i}-w_{i})m_{i+1}}
 \cdots 
e^{2\pi i (m_K-m_{K-1}) (w_{K-1} - u_{K-1} )}
\nn\\
&&\!\!\!\!\!\!\!\times
F_{2t}(w_{K-1}-u_{K-1})
F_{2v+\phi_K}(w_{K-1})F_{2v-\phi_K}(u_{K-1})\nn\\
&&\!\!\!\!\!\!\!\times
F_{-2v+\phi_1} (w_1+\xi_1)
F_{-2v+\phi_2} (w_2-w_1) \cdots 
e^{i\pi (w_i- w_{i-1})^2}   \cdots
F_{-2v+\phi_{K-1}} (w_{K-1}-w_{K-2}) 
\nn\\
&&\!\!\!\!\!\!\!\times
F_{-2v-\phi_1} (u_1-\xi_2)
F_{-2v-\phi_2} (u_2-u_1) \cdots e^{-i\pi (u_i- u_{i-1})^2} \cdots F_{-2v-\phi_{K-1}} (u_{K-1}-u_{K-2}) \,.
\nn\een
From this expression we see that the  axial mass deformation removes a bifundamental hyper on each tail and replaces it with  a BF coupling at level  $k=\pm 1$.

In this case the superpotential doesn't have the terms $ \M^{\pm(0,\ldots,0,1,\ldots1,1|1,1,\ldots,1,0\ldots,0)}$ which are dual to the cubic couplings  $P_i A Q_i+\tilde P_i \tilde A_i \tilde Q_i$  containing the integrated  out chirals.\\

\paragraph{$(1,K)$ and $(K,1)$ pq-5branes.}
By iterating the real mass deformation one can describe more general configurations. In particular the case where we integrate out all the  chirals  corresponding to moving to infinity in opposite directions the two stacks of $K$ $D5'$- halves on the two sides of the $NS5$, deforms the $(1_{NS5}, K_{D5'})$-web into a $(1,K)$ pq-5brane. Recall that a $D5'$ is a $(1,0)$ pq-5brane and an $NS5$ is a $(0,1)$ pq-5brane.

The low energy theory describing the configuration  $1_{NS5'}-1_{(1,K)5}-1_{NS5'}$ is then obtained by taking the limit 
$\phi_i=s \to \infty$ for $i=1,\cdots K$ in $\CT_{A,K}$ and it is given by a $U(1)^2$ quiver with a bifundamental, and $\pm K$ Chern-Simons units for each node.
The low energy description of the  S-dual set-up $1_{DS5'}-1_{(K,1)}-1_{D5'}$ can be  obtained by taking the same limit in $\CT_{C,K}$, which has the effect of removing  all the bifundamentals in the two tails and replacing  them by BF couplings. Since all the nodes in the tails have no matter we can easily perform all the Gaussian integrations and present the theory as a $U(1)^2$ quiver with a bifundamental, and $\pm 1/K$ Chern-Simons units for each node. The inversion of the Chern-Simon level is indeed what is expected by dualisation.

Actually since the low energy theory of the set-up  $1_{NS5'}-1_{(1,K)}-1_{NS5'}$ has only a bifundamental we 
 can perform all the integrations and find:
\ben
&&
\int dx_1 dx_2 e^{2\pi i (\xi_1 x_1+\xi_2 x_2)}  e^{i\pi K (x_2^2- x_1^2)}  ~F_{2t}(x_2-x_1)
= 2 e^{\tfrac{\pi i (\xi_1-\xi_2)^2}{K}}  F_{2t}(\xi_1+\xi_2)\,.
\een


\section{Abelian necklace quivers}\label{necklace}
\begin{figure}[h]
\centering
\includegraphics[width=4.5in]{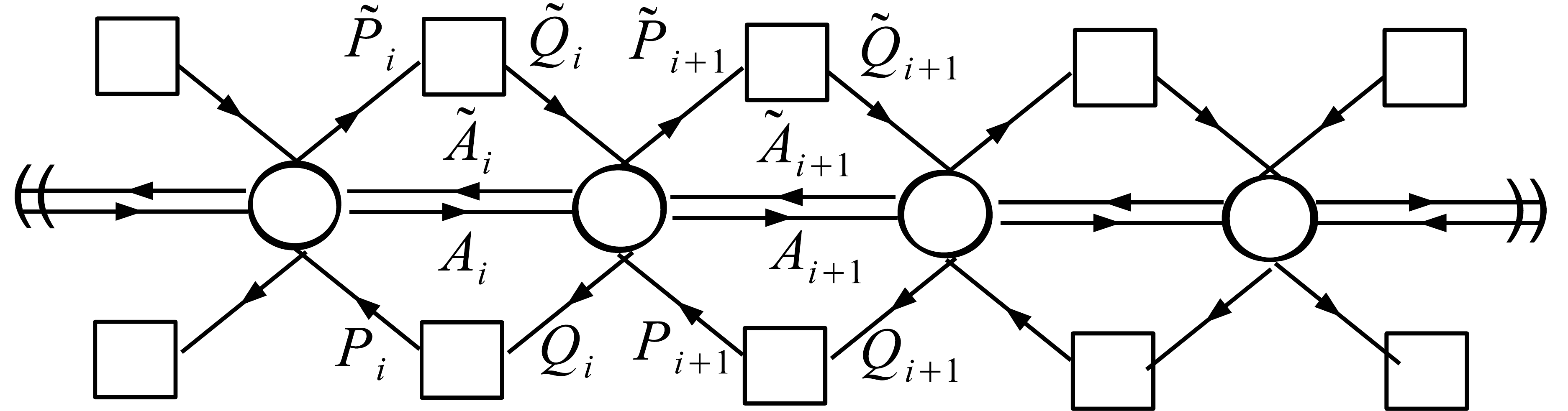}
\caption{ The necklace quiver. }
\label{neckpic}
\end{figure}

In this section we consider a circular, or necklace, quiver $U(1)^K$, with $K$ pairs of bifundamental chiral fields $A_i, \At_i$ and two pairs of chiral flavors $P_i, \Pt_i, Q_i, \Qt_i$ at each node as shown in Fig. \ref{neckpic}.
The superpotential is given by: 
\be \CW= \sum_{i=1}^K (\Phi_i A_i \At_i - \Phi_i A_{i+1} \At_{i+1} + A_iP_iQ_i + \At_i \Pt_i \Qt_i ) \,.\ee 
This quiver theory should be related to the low energy theory on  D3 branes stretched between $K$ pq-web of the basic form $(1_{D5'},1_{NS})\ldots - (1_{D5'},1_{NS}) - (1_{D5'},1_{NS}) - \ldots$. The $D3$ brane direction is a circle.\footnote{The quiver enjoys a $U(1)^{3K+1}$ global symmetry, while the brane setup should support $U(1)_t \times U(1)^{K}$ global symmetry (one $U(1)$ for each $(1_{D5'},1_{NS})$ pq-web). The discrepancy should be due to an accidental symmetry in the field theory description, similarly to what happens for the brane setup $NS-D5'-D5'-\ldots-D5'-NS$: the low energy $3d$ description is $U(1)$ with $k$ flavors and $\CW=0$. The global symmetry of the $U(1)$ gauge theory includes an $SU(k)^2$, while on the brane there is only one $SU(k)$ factor.}
The brane set up is invariant under the action of IIB S-duality and we expect the necklace quiver to be self mirror. We will show that this is indeed the case first by  implementing the piecewise dualisation on  the $S_b^3$ partition function, then by working out the map  of the chiral ring generators.

(Higher rank versions) of these type of quivers, known as Hand-Saw quivers, appear in a variety of contexts. For example they appear in connection with  Laumon spaces  and W-algebras \cite{Braverman:2010ef}, \cite{Nakajima:2011yq} providing a mathematical proof of a finite analog of the AGT correspondence in presence of surface operators.

Hand-Saw quivers play also  a central role in the work of \cite{Aganagic:2014oia} where they have three avatars.
They appear as the theories  on the vortices on the Higgs branch of  5d $\mathcal{N}=1$ $A_n$ quiver gauge theories.  They also describe the 5d gauge theories  when the  Coulomb branch parameters are tuned to special values and fluxes are turned on. The last incarnation is as $q$-deformed free field correlators of $q$-deformed $W_N$ primaries.

Here we focus on the self-mirror property of the abelian case leaving the discussion of the higher rank for a future work.

\subsection*{$S^3_b$ partition function}
The $S^3_b$ partition function with $K$ real masses $\xi_i$ for the topological symmetry, $K$ vector masses $\mu_i$ (with $\sum\mu_i=0$), $K$ axial masses $\phi_i$,  bifundamental mass $B$ and the usual masses $t$  and $2v=-t-\tfrac{iQ}{4}$ is given by:
\be
Z_{Necklace}=s_b(-2t)^K\prod_{i=1}^K\int dx_i e^{2\pi i x_i \xi_i}  F_{2t}(x_{i+1}-x_i+2B)F_{2v+\phi_i}(\mu_{i}+x_i+B)
F_{2v-\phi_{i+1}}(\mu_{i+1}+x_i-B)\,.
\label{necklaceeq}\ee
To find the piecewise  mirror we begin by  dualizing all fields by means of the pentagon identity  (\ref{fund}):
\ben
&&=\prod_{i=1}^K s_b(2v\pm \phi_i) \int dx_i  ds_i  dp_i dq_i    ~~F_{-2t}(s_i)F_{-2v-\phi_i}(p_i)F_{-2v+\phi_{i+1}}(q_i) \\
&&    e^{2\pi i x_i \xi_i} e^{-2\pi i s_i(x_{i+1}-x_ i+2B)}
 e^{-2\pi i p_i (\mu_{i}+x_i+B)}  e^{-2\pi i q_i (\mu_{i+1}+x_i-B)}\,.
 \een
Integrations over $x_i$ yield $\delta(\xi_i-q_i-p_i+s_i-s_{i-1})$ which we implement.
After shifting  $p_i \to p_i -s_{i-1}$ we find
\ben\nn
&&=\prod_{i=1}^K   e^{-2\pi i \xi_i  ( \mu_{i+1}-B)}  s_b(2v\pm \phi_i) \int      dp_i  e^{2\pi i p_i (\mu_{i+1}  -\mu_i-2B) } 
\\
&&
\times\int   ds_i~F_{-2t}(s_i)
F_{-2v-\phi_{i}}(p_i-s_{i-1})
F_{-2v+\phi_{i+1}}(\xi_i  +s_i-p_i   ) \,.
 \een 
We then use the new pentagon identity (\ref{newpent}) to perform the integrals $ds_i$.
Alternatively one can apply $K$-times the $\CT_C=\CT_A$  duality, which reduces to the  identity (\ref{Bide}), in   (\ref{necklaceeq}) and then implement the delta functions.

Now we shift $p_i \to p_i -\sum_{j=1}^{i-1} \xi_j$ so that for example $p_{i+1}-p_{i}\to p_{i+1}-p_{i}-\xi_i$ and obtain:
\ben\nn
&&=s_b(-2t)^K\prod_{i=1}^K   e^{-2\pi i \xi_i  ( \mu_{i+1}-B)}
  e^{2\pi i  ( - \sum_{j=1}^{i-1}\xi_j )(\mu_{i+1}-\mu_i-2B) }  
\int      dp_i  e^{2\pi i p_i (\mu_{i+1}  -\mu_i-2B) } 
\\
&&
F_{2t}( p_{1}-p_K  +\sum_{j=1}^K \xi_j )
(\prod_{j=1}^{K-1} F_{2t}( p_{j+1}-p_j))    
F_{2v+\phi_{i+1}}(\sum_{j=1}^i\xi_j     -p_i)
 F_{2v-\phi_{i+1}}(\sum_{j=1}^i\xi_j  -p_{i+1})\,.\nn\\
 \een 
Finally we  define $F=\sum_{j=1}^K \xi_j $,  shift $p_i \to p_i + i  \frac{F}{K}$ and, after few manipulations we arrive at the following result for the mirror dual necklace:\footnote{We omit a mirror-map invariant  prefactor containing background Chern-Simons terms.}
%
%
%
%
%
\ben\nn
&&=s_b(-2t)^K\prod_{i=1}^K
\int      dp_i  e^{2\pi i p_i (\mu_{i+1}  -\mu_i-2B) }  F_{2t}( p_{i+1}-p_i  +\tfrac{F}{K} ) 
\\
&&
 \times  F_{2v+\phi_{i+1}}  (\sum_{j=1}^i(\xi_j  -\tfrac{F}{K} ) +\tfrac{F}{2K}   -p_i) 
F_{2v-\phi_{i+1}}  (\sum_{j=1}^i(\xi_j  -\tfrac{F}{K} ) -\tfrac{F}{2K}   -p_{i+1}) =\widetilde Z_{Necklace}\,.
\label{dualnecklace}
 \een 
By comparing the   partition functions  (\ref{necklaceeq}) and  (\ref{dualnecklace}) it is easy to work out the mirror map:
\be
\label{mirmapne}
2B\to \tfrac{F}{K} \,, \qquad \xi_i \to  (\mu_{i+1}-\mu_i-2B) \,, \qquad  \mu_{i+1}\to -\sum_{j=1}^i(\xi_j  -\tfrac{F}{K} )
\,, \qquad  \phi_{i+1}\to -\phi_{i+1} \,, \qquad t \to t\,.
 \ee
As expected mirror symmetry acts by swapping the real masses associated to the topological symmetries with those associated to the flavor symmetries while axial masses change sign. This is similar to what happen in the mass deformed $TSU(N)$.

\subsection*{Chiral ring generators map}
The full moduli space is quite complicated and has many branches. Here we focus on the two extreme ones the Coulomb (where all charged fields are set to zero) and the Higgs branch (where monopole and singlets $\Phi_i$ are set to zero).

\paragraph{Higgs branch operators}
The Higgs branch component of chiral ring is generated by the $2K$ mesons $M_{Q,i}, M_{P,i} = Q_i\Qt_i, P_i\Pt_i$, and by $2K(K-1)$ length-$L$ ``extended mesons" operators of the form
\be  \qquad M^{i}_L = Q_{i}  \prod_{k=1}^L \tilde A_{i+k}  P_{i+L+1}  \,,\qquad \Mt^{i}_L = \Qt_{i} \prod_{k=1}^L  A_{i+k}  \Pt_{i+L+1} \,, \qquad L=0, \cdots , K-2\,,  \ee
where for $L=0$ we have $M^i_0=Q_i \Pt_{i+1}$ and $ \Mt^i_0=\tilde Q_i P_{i+1}$. 
These  operators start at node $i$ and end at node $i+L$ going in the right directions. $0 \leq L<K$, so the mesons don't wrap around the whole necklace. Actually also the longest mesons (for $L=K-1$)
\be  \qquad M^{i}_{K-1} = Q_{i}  \prod_{k=1}^{K-1} \tilde A_{i+k}  P_{i+K}  \,,\qquad \Mt^{i}_{K-1} = \Qt_{i} \prod_{k=1}^{K-1}  A_{i+k}  \Pt_{i+K} \,,  \ee
vanish  since $Q_iP_{i+K}=Q_iP_i$ and $\Qt_i\Pt_{i+K}=\Qt_i\Pt_i$ vanish on the Higgs branch where $\Phi_i=0$. 

We could also consider operators like $ \Qt_{i} \prod_{k=1}^L  A_{i+k}  Q_{i+L} $ but they are zero in the chiral ring because of $\CF$-terms.

There are also $3$ mesonic operators constructed  with bifundamental fields only:
\be M_B=A_i \At_i \qquad \CL= \prod_i A_i \qquad \CLt= \prod_i\At_i \,,\ee
satisfying $\CL \CLt = M_B^N$. 
There is only one meson $A_i\At_i$ because of the $\Phi_i$  $\CF$-terms.

In total we have $2K(K-1)+2K+3=2K^2+3$ mesonic generators, their scaling dimensions are easy to write down in terms of the scaling dimension $r$ of $P_i$:
\be R[\Mt^{i}_L]=R[M^{i}_L]= 2r + L (2-2r) \,,\ee
and
\be R[M_B] = 4-4r \qquad R[\CL, \CLt]=K(2-2r) \,.\ee
Notice that $R[M^{i}_1]=R[\Mt^{i}_1]=2$, so these $2K$ operators are marginal directions that could be added to the superpotential.
Also, there should be relations like
\be M^i_L \Mt^i_L = M_B^L M_{Q,i} M_{P,i+L} \,.\ee

\paragraph{Coulomb branch operators}
The Coulomb branch generators are the  $K$ singlets  $\Phi_i$ with scaling dimension $R[\Phi_i] = 4r-2$. 

Then we have $2K(K-1)$  basic monopole operators with topological charges given by strings of  
 $+1$'s or $-1$'s  starting at node $i$ and ending at node $i+L$ with $L<K$.
Their scaling dimension is given by: 
  \be R[\M^{\pm(0,\ldots,0,1_i,1_{i+1},\ldots,1_{i+L},0,0,\ldots,0)}]= 2 (1-R[A_i]) + 2(L+1)(1-R[P_i]) =
  2r+L(2-2r). \ee
Notice that the 2K monopoles with $L=1$ have scaling dimension 2 and can be added to the superpotential.  
  
Finally there are two basic  monopole operators wrapping the necklace, their topological charges are all $+1$ or all $-1$. Their R charge receives contribution only from the fundamental fields:
\be R[\M^{\pm(1,1,\ldots,1)}]=2K(1-R[P_i])=2K(1-r)\,. \ee

\paragraph{Mapping}

The previous discussion suggests the following map of operators under mirror symmetry.
Since 
\be R[\M^{\pm(1,1,\ldots,1)}]=R[\CL] = R[\CLt]\,, \ee
 the two wrapping monopoles are mapped to the two wrapping mesons $\CL$ and $\CLt$.
We can also see this by looking at effect of  the mirror map (\ref{mirmapne}) on
the real masses for of wrapping mesons
\be
K(\tfrac{iQ}{4}+t\pm 2B)\quad \to \quad K(\tfrac{iQ}{4}+t\pm\tfrac{F}{K})\,.
\ee

We then have  that \be R[M^{i}_L]=R[\Mt^{i}_L] = R[\M^{\pm(0,\ldots,0,1_i,1,\ldots,1_{i+L},0,0,\ldots,0)}] \,,\ee
which suggests that under mirror symmetry the $2K(K-1)$ basic monopoles are mapped to the extended mesons $M^i_L, \Mt^i_L$. 
For example we can see how the masses of the mesons  $\Qt_i\Pt_{i+1}, Q_iP_{i+1}$ are transformed by the mirror map (\ref{mirmapne}): 
\be
 \left(\tfrac{iQ}{2}+2v+ \tfrac{\phi_i-\phi_{i+1}}{2}  \pm(  \mu_{i+1}-\mu_i-2B)\right)
 \quad \to \quad 
 \left(\tfrac{iQ}{2}+2v- \tfrac{\phi_i-\phi_{i+1}}{2}  \pm  \xi_i\right)\,.
 \ee

Finally the $K$ singlets $\Phi_i$'s,  the bifundamental meson $M_B\simeq A_i\At_i$  and the $2K$ mesons $Q_i\Qt_i, P_i\Pt_i$ are mapped into themselves. 
Indeed they respectively have real masses equal to  $-2t$, $2t$
and $\tfrac{iQ}{2}+2v\pm\phi_i$ hence they  are left unchanged by the  mirror map (\ref{mirmapne}).\\

\begin{figure}[h]
\centering
\includegraphics[width=3.7in]{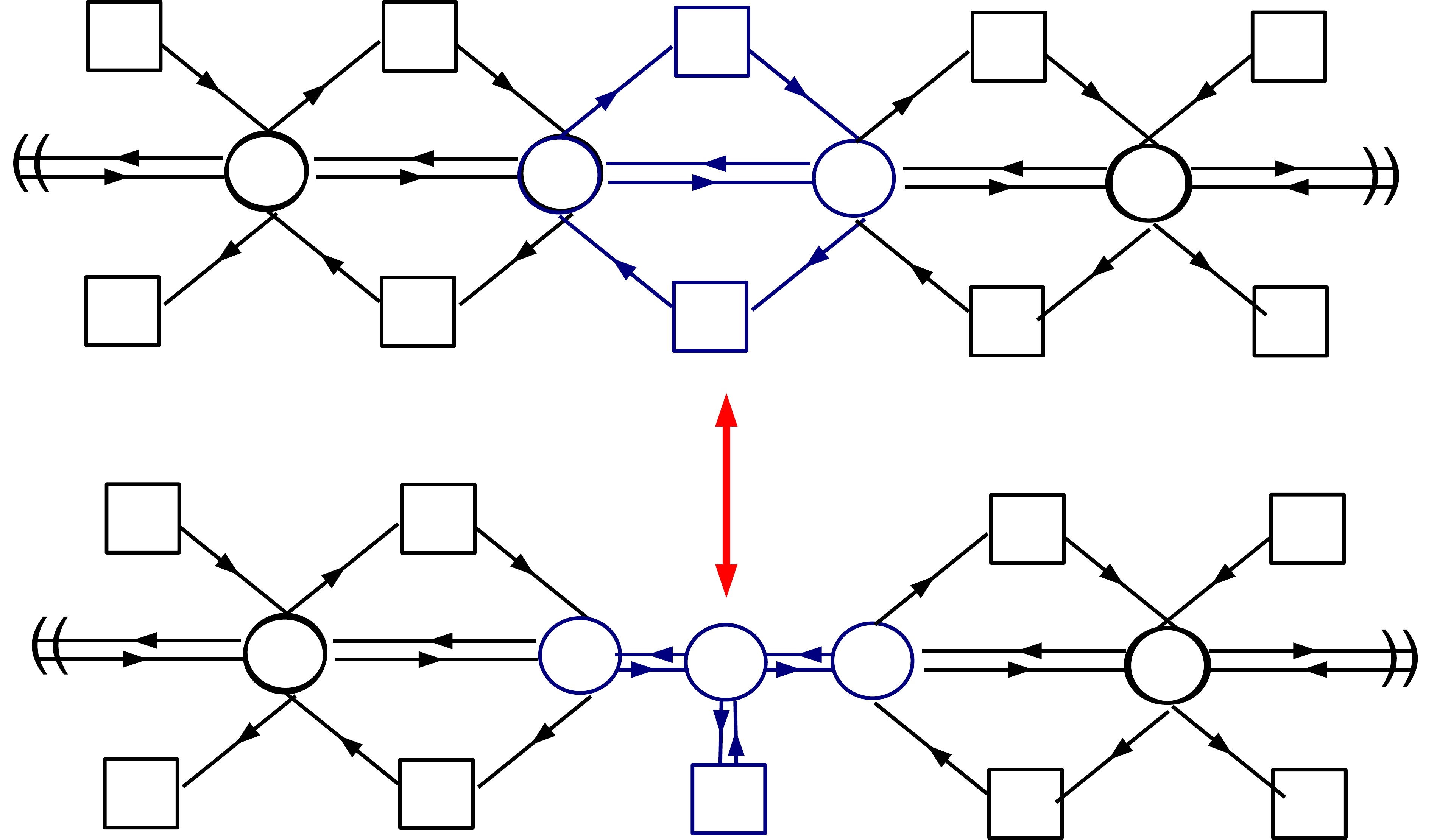}
\caption{We can apply the $\CT_C=\CT_B$ to generate many dual presentation of the necklace quiver.}
\label{magicneck}
\end{figure}

We close this section by noticing that while the  dual quiver we have been discussing in this section is obtain by applying $K$ times the identity $\CT_A=\CT_C$, we can find many dual presentations of the necklace quiver by applying
 the $\CT_C=\CT_B$ duality as shown Fig. \ref{magicneck}.

\acknowledgments
We are grateful to Francesco Benini, Stefano Cremonesi, Simone Giacomelli and Amihay Hanany for useful conversations.
 S.P. is partially  supported by the  ERC-STG grant 637844-HBQFTNCER. S.B. is partly supported by the INFN Research Projects GAST and ST$\&$FI and by PRIN "Geometria delle variet\`a algebriche".
 
\appendix
\section{Alternative proof of the 3d mirror symmetry $\CT_A \leftrightarrow \CT_C$}\label{LONG}
We want to show that $\CT_A$ and $\CT_C$ are dual, using dualities where the theories remain quivers at every step.

 The two $U(1)$ nodes in the quiver both have $2$ flavors, so we can use the mirror symmetry for $U(1)$ with 2 flavors reviewed earlier:
 \be U(1)_{p_i,\pt_i}, \CW=0 \leftrightarrow U(1)_{q_i,\qt_i}, \CW=\Phi_1 q_1 \qt_1 + \Phi_2 q_2 \qt_2\,. \ee 
 The mapping of the chiral ring generators is
 \be (p_1 \pt_1, p_2 \pt_2; p_1 \pt_2, p_2 \pt_1; \M^+, \M^-) \leftrightarrow (\Phi_1, \Phi_2; \M^+, \M^-; q_1\qt_2, q_2\qt_1 )\,.\ee
 
 Now let us flip the two mesons $p_1\pt_2$ and $p_2\pt_1$ of $U(1)/2, \CW=0$ and call the resulting theory $\CT$. Flipping accordingly on the right hand side, and calling the resulting theory $\CT'$, we obtain a new version of the duality:
  \be \CT: U(1)_{p_i,\pt_i}, \CW=\Phi_u p_1 \pt_2 + \Phi_d p_2 \pt_1 \leftrightarrow \CT': U(1)_{q_i,\qt_i}, \CW=\Phi_1 q_1 \qt_1 + \Phi_2 q_2 \qt_2 + \phi_u \M^+ + \phi_d \M^- \ee 
 The mapping of the chiral ring generators between $\CT$ and $\CT'$ reads
 \be (p_1 \pt_1,  p_2 \pt_2; \Phi_u, \Phi_d; \M^+, \M^- ) \leftrightarrow (\Phi_1, \Phi_2; \phi_u, \phi_d; q_1\qt_2, q_2\qt_1 )\ee
 
 Now we use Aharony duality for $U(1)$ with $2$ flavors in the form
  \be \label{AHA} U(1)_{q_1,q_2,\qt_1,\qt_2}, \CW=\phi_u \M^+ + \phi_d \M^- \leftrightarrow U(1)_{Q_1,Q_2,\Qt_1,\Qt_2}, \CW=\sum_{ij} \Phi_{ij} Q_i \Qt_j \,,\ee
  with 
 \be (q_1 \qt_1, q_2 \qt_2; q_1 \qt_2, q_2 \qt_1; \phi_u, \phi_d) \leftrightarrow (\Phi_{11}, \Phi_{22}; \Phi_{12}, \Phi_{21}; \M^+, \M^- )\,.\ee
 Flipping $q_1\qt_1$ and $q_2\qt_2$ on the l.h.s. of (\ref{AHA}) we obtain a duality between $\CT'$ and a third theory $U(1)/2$ with only the two off-diagonal mesons flipped, that we call $\CT''$:
  \be \CT': U(1)_{q_i,\qt_i}, \CW=\Phi_1 q_1 \qt_1 + \Phi_2 q_2 \qt_2 + \phi_u \M^+ + \phi_d \M^- \leftrightarrow 
                      \CT'': U(1)_{Q_i,\Qt_i}, \CW =  \phi_{1} Q_1 \Qt_2 +  \phi_{2} Q_2 \Qt_1\nn \,,\ee 
 \be (\Phi_1, \Phi_2; \phi_u, \phi_d; q_1 \qt_2, q_2 \qt_1) \leftrightarrow (Q_1\Qt_1, Q_2\Qt_2; \M^+, \M^- ; \phi_{1}, \phi_{2})\ee
 
 $\CT$ and $\CT''$ are the same theory, but following the mapping of the chiral rings we obtain a non trivial duality that exchanges the two chiral singlets $\Phi_u, \Phi_d$ with the two monopole operators $\M^+, \M^-$:
   \be \CT: U(1)_{p_i,\pt_i}, \CW=\Phi_u p_1 \pt_2 + \Phi_d p_2 \pt_1 \leftrightarrow   \CT'': U(1)_{Q_i,\Qt_i}, \CW =  \phi_{1} Q_1 \Qt_2 +  \phi_{2} Q_2 \Qt_1 \ee 
 \be (p_1 \pt_1,  p_2 \pt_2; \Phi_u, \Phi_d; \M^+, \M^- ) \leftrightarrow (Q_1\Qt_1, Q_2\Qt_2; \M^+, \M^-; \phi_{1}, \phi_{2}) \ee
 
  The symmetry we need to gauge, in order to go from $\CT$ to $\CT_A$, acts on the fields of $\CT$ $(p_1, \pt_1, p_2, \pt_2, \Phi_u, \Phi_d)$ with charges $(+1,-1,0,0,-1,+1)$, so it acts on the chiral ring generators $(p_1 \pt_1,  p_2 \pt_2; \Phi_u, \Phi_d; \M^+, \M^- )$ with charges $(0,0; +1, -1; 0, 0)$. The dual symmetry must have the same charges on the corresponding chiral ring generators: all the mesons in $\CT''$ have to be neutral, while the monopoles have to be charged. On $\CT''$ the symmetry is the topological symmetry, so gauging the symmetry means to ungauge the gauge symmetry of $\CT''$, and we get precisely $\CT_C$.
 
 We end up with a duality between $\CT_A = \cal{S} \cdot \CT$ and $\CT_C=   \cal{S} \cdot \CT''$.


\begin{thebibliography}{99}

\bibitem{Intriligator:1996ex} 
  K.~A.~Intriligator and N.~Seiberg,
  ``Mirror symmetry in three-dimensional gauge theories,''
  Phys.\ Lett.\ B {\bf 387}, 513 (1996)
  [hep-th/9607207].

\bibitem{Hanany:1996ie}
  A.~Hanany and E.~Witten,
  Nucl.\ Phys.\ B {\bf 492} (1997) 152
  doi:10.1016/S0550-3213(97)00157-0, 10.1016/S0550-3213(97)80030-2
  [hep-th/9611230].

\bibitem{Aharony:1997bx} 
  O.~Aharony, A.~Hanany, K.~A.~Intriligator, N.~Seiberg and M.~J.~Strassler,
  ``Aspects of N=2 supersymmetric gauge theories in three-dimensions,''
  Nucl.\ Phys.\ B {\bf 499}, 67 (1997)
  [hep-th/9703110].

\bibitem{deBoer:1997kr} 
  J.~de Boer, K.~Hori and Y.~Oz,
  ``Dynamics of N=2 supersymmetric gauge theories in three-dimensions,''
  Nucl.\ Phys.\ B {\bf 500}, 163 (1997)
  doi:10.1016/S0550-3213(97)00328-3
  [hep-th/9703100].
  
\bibitem{Aharony:1997ju} 
  O.~Aharony and A.~Hanany,
  ``Branes, superpotentials and superconformal fixed points,''
  Nucl.\ Phys.\ B {\bf 504}, 239 (1997)
  doi:10.1016/S0550-3213(97)00472-0
  [hep-th/9704170].

\bibitem{Brunner:1998jr} 
  I.~Brunner, A.~Hanany, A.~Karch and D.~Lust,
  ``Brane dynamics and chiral nonchiral transitions,''
  Nucl.\ Phys.\ B {\bf 528}, 197 (1998)
  doi:10.1016/S0550-3213(98)00318-6
  [hep-th/9801017].

\bibitem{Aharony:1997bh}
  O.~Aharony, A.~Hanany and B.~Kol,
  ``Webs of (p,q) five-branes, five-dimensional field theories and grid diagrams,''
  JHEP {\bf 9801} (1998) 002
  doi:10.1088/1126-6708/1998/01/002
  [hep-th/9710116].

\bibitem{Kapustin:1999ha} 
  A.~Kapustin and M.~J.~Strassler,
  ``On mirror symmetry in three-dimensional Abelian gauge theories,''
  JHEP {\bf 9904}, 021 (1999)
  doi:10.1088/1126-6708/1999/04/021
  [hep-th/9902033].

\bibitem{Pestun:2007rz}
  V.~Pestun,
  ``Localization of gauge theory on a four-sphere and supersymmetric Wilson loops,''
  Commun.\ Math.\ Phys.\  {\bf 313} (2012) 71
  doi:10.1007/s00220-012-1485-0
  [arXiv:0712.2824 [hep-th]].
  
\bibitem{Borokhov:2002ib} 
  V.~Borokhov, A.~Kapustin and X.~k.~Wu,
  ``Topological disorder operators in three-dimensional conformal field theory,''
  JHEP {\bf 0211}, 049 (2002)
  doi:10.1088/1126-6708/2002/11/049
  [hep-th/0206054].
  
\bibitem{Borokhov:2002cg} 
  V.~Borokhov, A.~Kapustin and X.~k.~Wu,
  ``Monopole operators and mirror symmetry in three-dimensions,''
  JHEP {\bf 0212}, 044 (2002)
  doi:10.1088/1126-6708/2002/12/044
  [hep-th/0207074].
  
\bibitem{Gaiotto:2008ak} 
  D.~Gaiotto and E.~Witten,
  ``S-Duality of Boundary Conditions In N=4 Super Yang-Mills Theory,''
  Adv.\ Theor.\ Math.\ Phys.\  {\bf 13}, no. 3, 721 (2009)
  doi:10.4310/ATMP.2009.v13.n3.a5
  [arXiv:0807.3720 [hep-th]].
  
\bibitem{Benna:2009xd} 
  M.~K.~Benna, I.~R.~Klebanov and T.~Klose,
  ``Charges of Monopole Operators in Chern-Simons Yang-Mills Theory,''
  JHEP {\bf 1001}, 110 (2010)
  doi:10.1007/JHEP01(2010)110
  [arXiv:0906.3008 [hep-th]].
  
\bibitem{Bashkirov:2010kz} 
  D.~Bashkirov and A.~Kapustin,
  ``Supersymmetry enhancement by monopole operators,''
  JHEP {\bf 1105}, 015 (2011)
  doi:10.1007/JHEP05(2011)015
  [arXiv:1007.4861 [hep-th]].
  
\bibitem{Imamura:2011su}
  Y.~Imamura and S.~Yokoyama,
  ``Index for three dimensional superconformal field theories with general R-charge assignments,''
  JHEP {\bf 1104} (2011) 007
  doi:10.1007/JHEP04(2011)007
  [arXiv:1101.0557 [hep-th]].
  
  
 
\bibitem{Cremonesi:2013lqa} 
  S.~Cremonesi, A.~Hanany and A.~Zaffaroni,
  ``Monopole operators and Hilbert series of Coulomb branches of $3d$  $\mathcal{N} = 4$ gauge theories,''
  JHEP {\bf 1401}, 005 (2014)
  doi:10.1007/JHEP01(2014)005
  [arXiv:1309.2657 [hep-th]].
  
\bibitem{Cremonesi:2014kwa} 
  S.~Cremonesi, A.~Hanany, N.~Mekareeya and A.~Zaffaroni,
  ``Coulomb branch Hilbert series and Hall-Littlewood polynomials,''
  JHEP {\bf 1409}, 178 (2014)
  doi:10.1007/JHEP09(2014)178
  [arXiv:1403.0585 [hep-th]].
  
\bibitem{Cremonesi:2014vla} 
  S.~Cremonesi, A.~Hanany, N.~Mekareeya and A.~Zaffaroni,
  ``Coulomb branch Hilbert series and Three Dimensional Sicilian Theories,''
  JHEP {\bf 1409}, 185 (2014)
  doi:10.1007/JHEP09(2014)185
  [arXiv:1403.2384 [hep-th]].
  

\bibitem{Polyakov:1976fu} 
  A.~M.~Polyakov,
  ``Quark Confinement and Topology of Gauge Groups,''
  Nucl.\ Phys.\ B {\bf 120}, 429 (1977).
  doi:10.1016/0550-3213(77)90086-4
    
\bibitem{Affleck:1982as} 
  I.~Affleck, J.~A.~Harvey and E.~Witten,
  ``Instantons and (Super)Symmetry Breaking in (2+1)-Dimensions,''
  Nucl.\ Phys.\ B {\bf 206}, 413 (1982).
  doi:10.1016/0550-3213(82)90277-2
  
\bibitem{Seiberg:1996nz} 
  N.~Seiberg and E.~Witten,
  ``Gauge dynamics and compactification to three-dimensions,''
  In *Saclay 1996, The mathematical beauty of physics* 333-366
  [hep-th/9607163].

  
\bibitem{Aharony:2013dha} 
  O.~Aharony, S.~S.~Razamat, N.~Seiberg and B.~Willett,
  ``3d dualities from 4d dualities,''
  JHEP {\bf 1307}, 149 (2013)
  doi:10.1007/JHEP07(2013)149
  [arXiv:1305.3924 [hep-th]].
  
\bibitem{Collinucci:2016hpz} 
  A.~Collinucci, S.~Giacomelli, R.~Savelli and R.~Valandro,
  ``T-branes through 3d mirror symmetry,''
  arXiv:1603.00062 [hep-th].
  

\bibitem{Dimofte:2012pd} 
  T.~Dimofte and D.~Gaiotto,
  ``An E7 Surprise,''
  JHEP {\bf 1210}, 129 (2012)
  doi:10.1007/JHEP10(2012)129
  [arXiv:1209.1404 [hep-th]].

\bibitem{Kapustin:2009kz} 
  A.~Kapustin, B.~Willett and I.~Yaakov,
  ``Exact Results for Wilson Loops in Superconformal Chern-Simons Theories with Matter,''
  JHEP {\bf 1003}, 089 (2010)
  doi:10.1007/JHEP03(2010)089
  [arXiv:0909.4559 [hep-th]].

\bibitem{Dimofte:2011ju}
  T.~Dimofte, D.~Gaiotto and S.~Gukov,
  ``Gauge Theories Labelled by Three-Manifolds,''
  Commun.\ Math.\ Phys.\  {\bf 325} (2014) 367
  doi:10.1007/s00220-013-1863-2
  [arXiv:1108.4389 [hep-th]].

\bibitem{Hama:2010av}
  N.~Hama, K.~Hosomichi and S.~Lee,
  ``Notes on SUSY Gauge Theories on Three-Sphere,''
  JHEP {\bf 1103} (2011) 127
  doi:10.1007/JHEP03(2011)127
  [arXiv:1012.3512 [hep-th]].

\bibitem{Jafferis:2010un}
  D.~L.~Jafferis,
  ``The Exact Superconformal R-Symmetry Extremizes Z,''
  JHEP {\bf 1205} (2012) 159
  doi:10.1007/JHEP05(2012)159
  [arXiv:1012.3210 [hep-th]].


\bibitem{Dimofte:2011py}
  T.~Dimofte, D.~Gaiotto and S.~Gukov,
  ``3-Manifolds and 3d Indices,''
  Adv.\ Theor.\ Math.\ Phys.\  {\bf 17} (2013) no.5,  975
  doi:10.4310/ATMP.2013.v17.n5.a3
  [arXiv:1112.5179 [hep-th]].





\bibitem{Aharony:1997gp} 
  O.~Aharony,
  ``IR duality in d = 3 N=2 supersymmetric USp(2N(c)) and U(N(c)) gauge theories,''
  Phys.\ Lett.\ B {\bf 404}, 71 (1997)
  doi:10.1016/S0370-2693(97)00530-3
  [hep-th/9703215].
  
  
  

 

\bibitem{Volkov2003}
  R.~Kashaev, F.~Luo and G.~Vartanov,
  ``Noncommutative Hypergeometry,''
  Communications in Mathematical Physics
September 2005, Volume 258, Issue 2, pp 257-273.
 arXiv:math/0312084 [math.QA]. 
  
\bibitem{Kashaev:2012cz}
  R.~Kashaev, F.~Luo and G.~Vartanov,
  ``A TQFT of Turaev-Viro type on shaped triangulations,''
  arXiv:1210.8393 [math.QA].



\bibitem{spiri}
V. P.~ Spiridonov,
  ``On the elliptic beta function,''
Uspekhi Mat. Nauk 56 (1) (2001) 181Ð182 (Russian Math.
Surveys 56 (1) (2001) 185Ð186)



\bibitem{Gahramanov:2013rda}
  I.~Gahramanov and H.~Rosengren,
  JHEP {\bf 1311} (2013) 128
  doi:10.1007/JHEP11(2013)128
  [arXiv:1309.2195 [hep-th]].


\bibitem{toappear1}
 Work in progress.

\bibitem{Gaiotto:2014ina}
  D.~Gaiotto and H.~C.~Kim,
  ``Surface defects and instanton partition functions,''
  arXiv:1412.2781 [hep-th].


\bibitem{Benvenuti:2006qr} 
  S.~Benvenuti, B.~Feng, A.~Hanany and Y.~H.~He,
  ``Counting BPS Operators in Gauge Theories: Quivers, Syzygies and Plethystics,''
  JHEP {\bf 0711}, 050 (2007)
  doi:10.1088/1126-6708/2007/11/050
  [hep-th/0608050].
  





  
\bibitem{Hanany:2015via} 
  A.~Hanany, C.~Hwang, H.~Kim, J.~Park and R.~K.~Seong,
  ``Hilbert Series for Theories with Aharony Duals,''
  JHEP {\bf 1511}, 132 (2015)
  Addendum: [JHEP {\bf 1604}, 064 (2016)]
  doi:10.1007/JHEP11(2015)132, 10.1007/JHEP04(2016)064
  [arXiv:1505.02160 [hep-th]].
  
\bibitem{Cremonesi:2015dja} 
  S.~Cremonesi,
  ``The Hilbert series of 3d ${\boldsymbol{\mathcal{N}}}=2$ Yang-Mills theories with vectorlike matter,''
  J.\ Phys.\ A {\bf 48}, no. 45, 455401 (2015)
  doi:10.1088/1751-8113/48/45/455401
  [arXiv:1505.02409 [hep-th]].
  
  


\bibitem{Cremonesi:2010ae}
  S.~Cremonesi,
  ``Type IIB construction of flavored ABJ(M) and fractional M2 branes,''
  JHEP {\bf 1101} (2011) 076
  doi:10.1007/JHEP01(2011)076
  [arXiv:1007.4562 [hep-th]].

  
    
\bibitem{Braverman:2010ef}
  A.~Braverman, B.~Feigin, M.~Finkelberg and L.~Rybnikov,
  ``A Finite analog of the AGT relation I: F inite $W$-algebras and quasimaps' spaces,''
  Commun.\ Math.\ Phys.\  {\bf 308} (2011) 457
  doi:10.1007/s00220-011-1300-3
  [arXiv:1008.3655 [math.AG]].
 
\bibitem{Nakajima:2011yq}
  H.~Nakajima,
  ``Handsaw quiver varieties and finite W-algebras,''
  arXiv:1107.5073 [math.QA].

  
\bibitem{Aganagic:2014oia}
  M.~Aganagic, N.~Haouzi and S.~Shakirov,
  ``$A_n$-Triality,''
  arXiv:1403.3657 [hep-th].

  
  \end{thebibliography}
\end{document}